\documentclass[usenatbib]{mn2e}

\usepackage{aas_macros}
\usepackage{amsmath}
\usepackage{amssymb}
\usepackage[colorlinks=true,breaklinks=true,citecolor=blue]{hyperref} % comment removed
\usepackage{xspace}
\usepackage{color}
\usepackage{graphicx}
\usepackage{txfonts} % comment removed
\usepackage{mathtools} % comment removed
\usepackage{tensor}

\newcommand{\eg}{\mbox{\textit{e.\,g.},}\xspace} % comment removed
\newcommand{\ie}{\mbox{\textit{i.\,e.},}\xspace}

\newcommand{\Planck}{\textit{Planck}}

\renewcommand{\vec}[1]{\bmath{#1}} % comment removed
\newcommand{\mat}[1]{\mathbfss{#1}}

\definecolor{blue}{rgb}{0,0,1} % comment removed
\definecolor{yellow}{rgb}{1,0.84,0}
\definecolor{red}{rgb}{1,0,0}
\definecolor{mypink1}{rgb}{0.858, 0.188, 0.478}
\definecolor{darkgreen}{rgb}{0,0.5,0}
\definecolor{orange}{rgb}{1.0,0.4,0}
\newcommand{\kkr}[1]{\textcolor{black}{#1}} % comment removed

\title[Spin-SILC: CMB polarisation component separation]{Spin-SILC: CMB polarisation component separation with spin wavelets}
\author[K.~K.~Rogers et al.]
{Keir K.~Rogers,$^{1}$\thanks{E-mail: keir.rogers.14@ucl.ac.uk} Hiranya V.~Peiris,$^{1}$ Boris Leistedt,$^{2,1}$ Jason D.~McEwen$^{3}$
\newauthor and Andrew Pontzen$^{1}$
\\
$^1$Department of Physics \& Astronomy, University College London, Gower Street, London WC1E 6BT, UK\\
$^2$Department of Physics, New York University, 4 Washington Place, New York, NY 10003, USA\\
$^3$Mullard Space Science Laboratory, University College London, Surrey RH5 6NT, UK}

\begin{document}
\maketitle
\begin{abstract}
We present Spin-SILC, a new foreground component separation method that accurately extracts the cosmic microwave background (CMB) polarisation \(E\) and \(B\) modes from raw multifrequency Stokes \(Q\) and \(U\) measurements of the microwave sky. Spin-SILC is an internal linear combination method that uses spin wavelets to analyse the spin-2 polarisation signal \(P = Q + iU\). The wavelets are additionally directional (non-axisymmetric). This allows different morphologies of signals to be separated and therefore the cleaning algorithm is localised using an additional domain of information. The advantage of spin wavelets over standard scalar wavelets is to simultaneously and self-consistently probe scales and directions in the polarisation signal \(P = Q + iU\) and in the underlying \(E\) and \(B\) modes, therefore providing the ability to perform component separation and \(E\)-\(B\) decomposition concurrently for the first time. We test Spin-SILC on full-mission {\Planck} simulations and data and show the capacity to correctly recover the underlying cosmological \(E\) and \(B\) modes. We also demonstrate a strong consistency of our CMB maps with those derived from existing component separation methods. Spin-SILC can be combined with the pseudo- and pure \(E\)-\(B\) spin wavelet estimators presented in a companion paper to reliably extract the cosmological signal in the presence of complicated sky cuts and noise. Therefore, it will provide a computationally-efficient method to accurately extract the CMB \(E\) and \(B\) modes for future polarisation experiments.
\end{abstract}
% comment removed

\section{Introduction}
\label{sec:intro}

The polarisation of the cosmic microwave background (CMB) is a powerful cosmological observable, providing deep insights into the physics of the early universe. The decomposition of the linear polarisation into curl-free (\(E\) mode) and divergence-free (\(B\) mode) components allows the detection of tensor perturbations to the metric. Specifically, a non-zero \(BB\) power spectrum on degree scales would support the existence of a stochastic background of gravitational waves predicted by inflationary theory \citep{1997PhRvL..78.2058K,1997PhRvL..78.2054S}. Accurate measurement of \(B\) mode polarisation on arcminute scales also gives strong constraints on the neutrino sector via the weak gravitational lensing of CMB \(E\) modes \citep{1998PhRvD..58b3003Z}. There are numerous existing and planned ground-based, balloon-borne and satellite experiments designed to precisely measure CMB polarisation (see \eg \citealt{2015arXiv150906770E} for a recent forecast on the cosmological constraining power of current and upcoming missions). However, as in measuring the temperature \(T\) anisotropies of the CMB, the polarised background needs to be separated from instrumental noise and signals due to astronomical foregrounds (in particular, synchrotron radiation and thermal radiation from Galactic dust). This foreground component separation is more difficult compared with the case of CMB temperature, due to the relative strength and morphological complexity of polarised foregrounds, which are poorly understood.

Foreground component separation has been performed in numerous ways but, on real observational data, always by removing foreground contamination from scalar signals. For example, in the polarised setting, foreground contamination is removed from the Stokes \(Q\) and \(U\) or from \(E\) and \(B\) mode maps by treating \(Q\) and \(U\) or \(E\) and \(B\) as independent scalar fields. We presented a thorough discussion of blind and non-blind component separation methods in \citet{2016arXiv160101322R} (see also \eg \citealt{2009A&A...493..835D,2013A&A...550A..73B} for reviews). In this work, we highlight only the four component separation methods employed in \citet{2015arXiv150205956P}. Commander \citep{2006ApJ...641..665E,2008ApJ...676...10E} and SEVEM \citep{2003MNRAS.345.1101M,2008A&A...491..597L,2012MNRAS.420.2162F} operate on the \(Q\) and \(U\) maps, while NILC \citep{2009A&A...493..835D} and SMICA \citep{2008arXiv0803.1814C} operate on the \(E\) and \(B\) mode maps.

Recently, \citet{2016arXiv160101515F} explored an extension of the Internal Linear Combination (ILC) method to act fully on the spin-2 signal formed by the \(Q\) and \(U\) maps and applied their method to simulations. In general, the ILC method estimates the CMB as a weighted sum of maps of the sky at different microwave frequencies. The weights are constrained to conserve the CMB signal but minimise foreground and noise residuals by minimising the variance of the output map. The weights can be localised in various domains, but most usefully in wavelet space \citep[\eg][]{2016arXiv160101322R}, which allows the weights to vary simultaneously with position on the sky and harmonic scale. \citet{2016arXiv160101515F} minimised the covariant quantity \(\langle {|P|}^2 \rangle\), where \(P = Q + iU\), in map space. Consequently, they do not consider any harmonic localisation.

In this work, we introduce the Spin, Scale-discretised, directional wavelet ILC or Spin-SILC. This is an extension of the SILC method we introduced in \citet{2016arXiv160101322R}, where to analyse spin signals such as CMB polarisation we now use spin scale-discretised wavelets, the complete construction of which is presented in \citet{mcewen:s2let_spin} \citep[see also][]{2015arXiv150203120L, 2014IAUS..306...64M}. Wavelets are functions that are localised in both real and harmonic space and, in particular, scale-discretised wavelets satisfy excellent localisation properties \citep{mcewen:s2let_localisation}. In the scalar SILC method, we use directional scale-discretised wavelets \citep{2008MNRAS.388..770W,2013SPIE.8858E..0IM,mcewen:s2let_spin}. Directional wavelets are spatially and harmonically localised and additionally ``directionally-localised,'' \ie the spatial kernels are non-axisymmetric and can be rotated to pick out a preferred direction on the surface of the sphere. \citet{2016arXiv160101322R} gives an introductory summary of directional wavelets. The spin wavelets we use are still spatially, harmonically and directionally localised but are now constructed in the space of spin spherical harmonics. When spin wavelets are convolved with spin signals defined on the sphere, the output wavelet coefficients isolate signal structure of different scale and orientation, while maintaining the spatial information. The spin can in general be arbitrary, but since we are interested in analysing the spin-2 signal \(P = Q + iU\) we adopt spin-2 wavelets. By the particular construction of the spin wavelets, the complex spin-2 wavelet coefficients can be separated (by their real and imaginary parts) into scalar wavelet coefficients of the \(E\) and \(B\) fields, where the scalar wavelet coefficients correspond to a scalar wavelet that is a spin-lowered version of the original spin-2 wavelet \citep{mcewen:s2let_spin}. Hence, the spin-2 wavelet transform at the heart of Spin-SILC performs \(E\)-\(B\) decomposition from input \(Q\) and \(U\) maps. The ILC method is then applied to the complex wavelet coefficients, with complex weights, and jointly minimises the variance of the reconstructed \(E\) and \(B\) fields. Moreover, the weights vary spatially, harmonically and according to different orientations, fine-tuning the cleaning algorithm to remove foreground and noise contamination.

It follows that Spin-SILC introduces two main novelties to CMB polarisation component separation. Firstly, the use of spin scale-discretised wavelets allows the full analysis of the polarisation spin signal \(P\). By their construction, we can then perform component separation and \(E\)-\(B\) decomposition simultaneously and self-consistently. Secondly, the use of directional wavelets allows the additional flexibility to localise the foreground removal according to the morphological structure of the CMB and the foregrounds.

There is a third novel attribute to Spin-SILC of interest to future polarisation observations. Although in this work, we have tested Spin-SILC on the full-sky multifrequency maps provided by the Planck Collaboration, these frequency maps are dominated by instrumental noise and hence so are also our estimates of the CMB polarisation. Future polarisation measurements will have high signal-to-noise, but will usually cover only a fraction of the sky. However, \(E\)-\(B\) decomposition (from the measured \(Q\) and \(U\) modes) on the cut-sky is not uniquely defined unlike the full-sky case. This leads to leaking or mixing between the \(E\) and \(B\) modes. This is of particular concern in extracting the \(B\) field since the \(E\) field is orders of magnitude larger. As presented in a companion paper \citep{ebsep}, the spin scale-discretised wavelets we use can be employed to construct pure estimators of the masked \(E\) and \(B\) modes (pure \(E\) (\(B\)) modes are orthogonal to all \(B\) (\(E\)) modes on the partial sky, respectively). This builds on the work of \citet{2002PhRvD..65b3505L,2003PhRvD..67b3501B,2007PhRvD..76d3001S,2012PhRvD..86g6005G,2013PhRvD..88b3524F} (see in particular \citealt{2003PhRvD..67b3501B} for a discussion of pure modes at the map level). This only requires calculating additional wavelet transforms of the input data subject to a suitably apodised mask. One of the main advantages of this approach is the possibility of optimising the mask as a function of scale and direction, therefore yielding a more efficient cancellation of the systematic \(E\)-\(B\) mixing due to masking \citep{ebsep}. Hence, Spin-SILC can produce accurate estimates of the cosmological \(E\) and \(B\) fields, even on the cut-sky, in conjunction with the \(E\)-\(B\) estimators presented in \citet{ebsep}. More details of how Spin-SILC can operate on partial sky observations are given in \S~\ref{sec:cut_sky}.

We provide an introduction to spin scale-discretised wavelets in \S~\ref{sec:spin_wav}. In \S~\ref{sec:method}, the Spin-SILC algorithm is explained in detail. We test the method on {\Planck} simulations in \S~\ref{sec:simulations} and {\Planck} data in \S~\ref{sec:data}. In \S~\ref{sec:comp}, we compare our method to previous component-separation methods. We discuss the results in \S~\ref{sec:discussion} and conclude in \S~\ref{sec:concs}.

\begin{figure}
\includegraphics[width=\columnwidth]{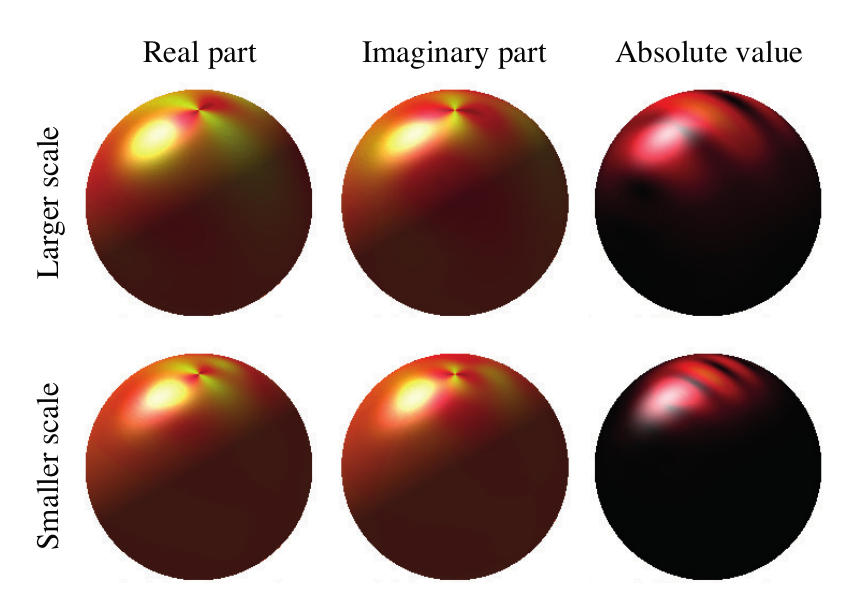}
\caption{The spatial localisation on the sphere of spin, directional, scale-discretised wavelets. The top row shows larger scale wavelets than the bottom row. The left column shows the real part of the wavelet, the middle column shows the imaginary part of the wavelet and the right column shows the absolute value of the wavelet. The number of directions per wavelet scale \(N = 5\). Therefore, for complete reconstruction at each scale, the above wavelets would be complemented by four more wavelets of the same size but of a different orientation on the sphere. The spin number \(s = 2\), which is what is required for the analysis of Stokes \(Q\) and \(U\) modes. This figure is adapted from \citet{mcewen:s2let_spin}.}
\label{fig:wavelets_spatial}
\end{figure}

\begin{figure}
\includegraphics[width=\columnwidth]{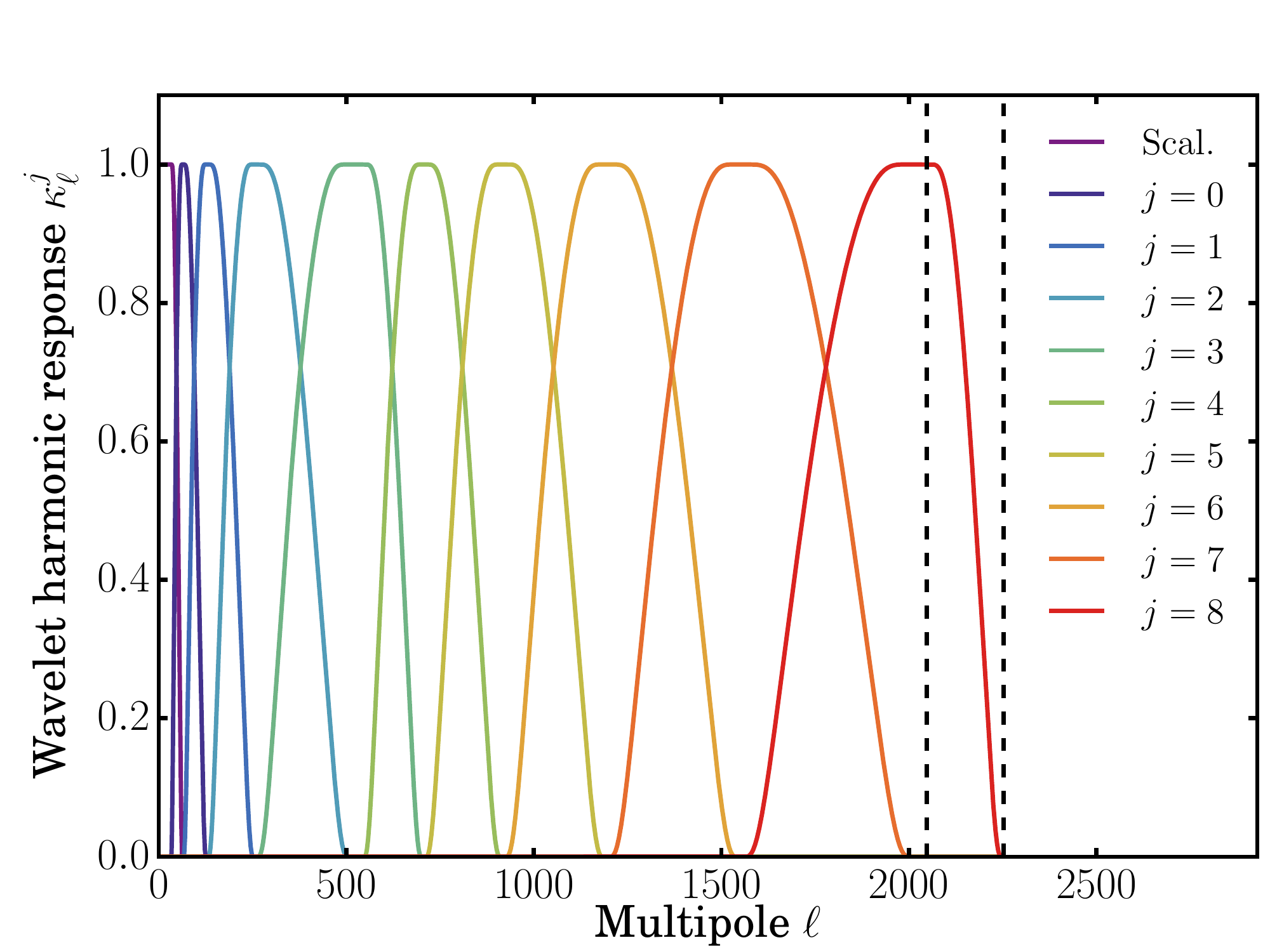}
\caption{The harmonic localisation of the spin wavelets used in this work (\(\kappa^j_\ell\) as defined in Eq.~\eqref{eq:wavelets}), where \(j\) specifies the wavelet scale. Increasing \(j\) corresponds to a smaller wavelet kernel and so a multipole range on smaller scales (\ie larger multipoles \(\ell\)). The largest wavelet scale (\(\mathrm{Scal.}\)) is the scaling function (\S~\ref{sec:analysis}). This choice of wavelets deliberately ensures exact reconstruction only for \(\ell \leq 2048\). The tapering of the smallest wavelet for \(2048 < \ell \leq 2253\) suppresses the smallest-scale power within the algorithm. The band-limits of the above wavelets are given in Table \ref{tab:wav_lims}.}
\label{fig:wavelets_harmonic}
\end{figure}

\section{Spin wavelets}
\label{sec:spin_wav}

\begin{table}\centering
\caption{The harmonic band-limits \([\ell^j_{\mathrm{min}},\ell^j_{\mathrm{max}}]\) of the spin wavelets used in this work. \(\ell^j_{\mathrm{peak}}\) is the multipole at which each wavelet has its maximum response. The final column shows the number of equiangular samples per wavelet coefficient map \(N^j_\mathrm{samp}\).}
\label{tab:wav_lims}
\begin{tabular}{ccccc}
\hline
Wavelet scale \(j\) & \(\ell^j_{\mathrm{min}}\) & \(\ell^j_{\mathrm{peak}}\) & \(\ell^j_{\mathrm{max}}\) & \(N^j_\mathrm{samp}\) \\
\hline
Scal. & 0 & 64 & 64 & 8,385 \\
0 & 32 & 64 & 128 & 33,153 \\
1 & 64 & 128 & 256 & 131,841 \\
2 & 128 & 256 & 512 & 525,825 \\
3 & 256 & 512 & 706 & 998,991 \\
4 & 542 & 705 & 918 & 1,688,203 \\
5 & 705 & 917 & 1193 & 2,850,078 \\
6 & 917 & 1192 & 1551 & 4,815,856 \\
7 & 1192 & 1550 & 2015 & 8,126,496 \\
8 & 1550 & 2015 & 2253 & 10,158,778 \\
\hline
\end{tabular}
\end{table}

\begin{figure*}
\iffalse
\begin{tabular}{cc}
\includegraphics[width=0.75\columnwidth]{cmb_orig_Qmap.pdf} \includegraphics[width=0.25\columnwidth]{cmb_orig_Qmap_zoom.pdf} & \includegraphics[width=0.75\columnwidth]{cmb_orig_Umap.pdf} \includegraphics[width=0.25\columnwidth]{cmb_orig_Umap_zoom.pdf} \\
\hline
Real part of spin-2 wavelets & Imaginary part of spin-2 wavelets \\
 & \\
\includegraphics[width=\columnwidth]{spin_wavelets_real.png} & \includegraphics[width=\columnwidth]{spin_wavelets_imag.png} \\
\hline
Real part of spin-2 wavelet transform of \(P = Q + iU\) & Imaginary part of spin-2 wavelet transform of \(P = Q + iU\) \\
\(=\) Scalar wavelet transform of \(E\) map & \(=\) Scalar wavelet transform of \(B\) map \\
 & \\
\includegraphics[width=\columnwidth]{cmb_decomp_scalar_E.png} & \includegraphics[width=\columnwidth]{cmb_decomp_scalar_B.png} \\
\hline
\includegraphics[width=0.75\columnwidth]{cmb_recon_Emap.pdf} \includegraphics[width=0.25\columnwidth]{cmb_recon_Emap_zoom.pdf} & \includegraphics[width=0.75\columnwidth]{cmb_recon_Bmap.pdf} \includegraphics[width=0.25\columnwidth]{cmb_recon_Bmap_zoom.pdf}
\end{tabular}
\fi
\includegraphics[width=\textwidth]{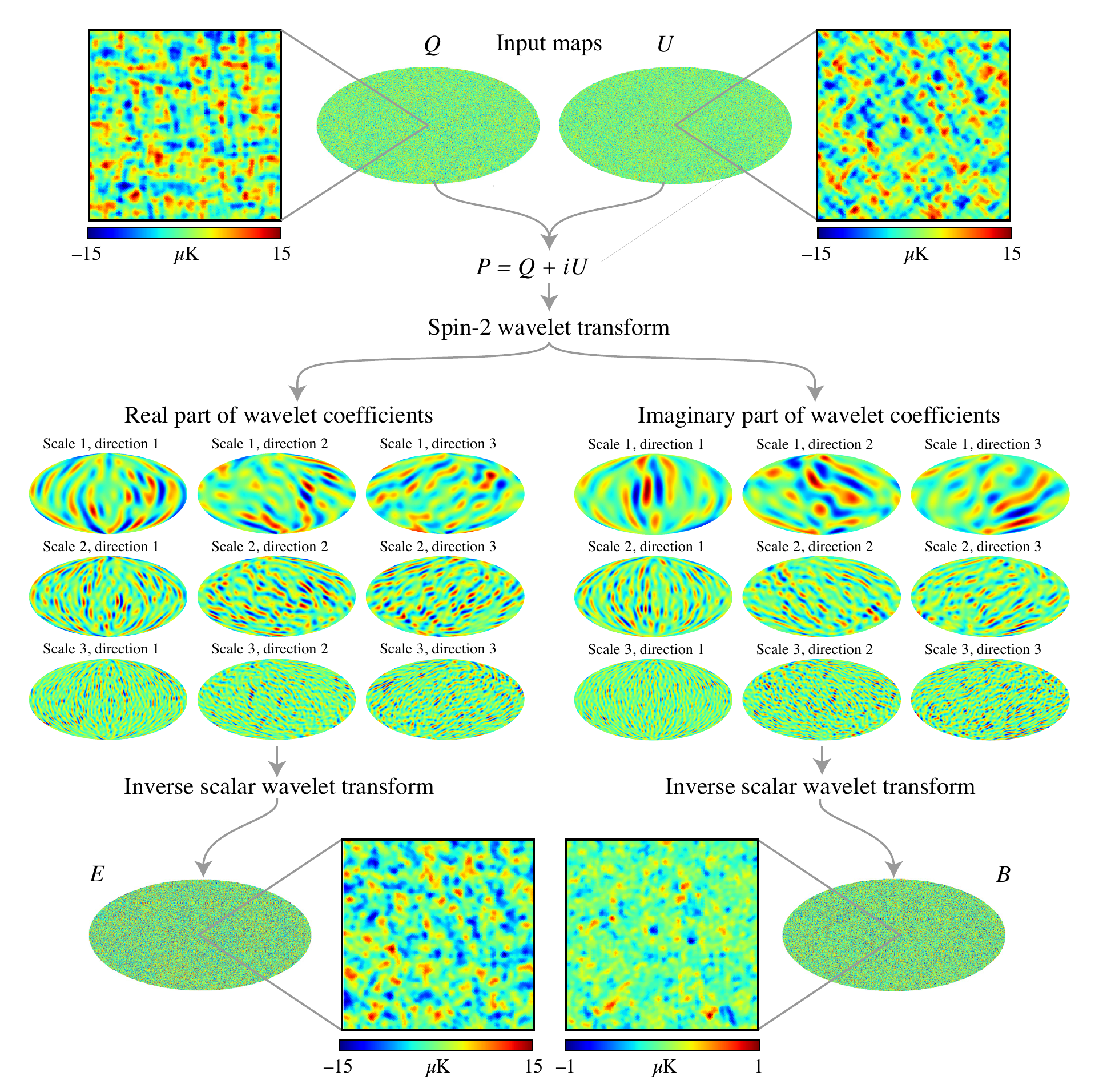}
\caption{An illustration of the spin, directional wavelet decomposition of the CMB Stokes \(Q\) and \(U\) maps and the \(E\) - \(B\) separation that automatically occurs as a consequence. The top row shows example input \(Q\) and \(U\) maps, simulated with lensed scalar perturbations, with zoomed regions to show structure in the fields. The middle row shows the real and imaginary parts of the spin, directional wavelet coefficient maps, formed by the spin-2 wavelet transform of \(P = Q + iU\). The ILC algorithm acts on such wavelet coefficients (calculated for multiple polarisation channels) and produces clean wavelet coefficients of the CMB polarisation. By the construction of the wavelets, the real and imaginary parts are respectively equal to scalar wavelet transforms of the \(E\) and \(B\) fields (with a different scalar wavelet). The bottom row shows the output \(E\) and \(B\) maps, also with zoomed regions, formed respectively by inverse scalar wavelet transforms of the real and imaginary parts of the wavelet coefficient maps. In our Spin-SILC analysis we include wavelets on smaller scales than those used in the simple demonstration shown above. }
\label{fig:cmb_decomp}
\end{figure*}

Spin, directional, scale-discretised wavelets on the sphere that support exact reconstruction have been constructed in \citet{mcewen:s2let_spin} (and discussed briefly in \citealt{2014IAUS..306...64M,2015arXiv150203120L}). These are an extension of the scalar, directional wavelets developed in \citet{2008MNRAS.388..770W,2013SPIE.8858E..0IM}, which are used in the scalar version of the SILC method for the analysis of CMB temperature anistropies \citep{2016arXiv160101322R}. They maintain the properties of spatial, harmonic and directional localisation, but can now additionally analyse spin fields by being constructed on the basis of spin spherical harmonics. In particular, spin-2 wavelets can be convolved with the spin-2 field \(P = Q + iU\), where \(Q\) and \(U\) are the Stokes parameters of the CMB's linear polarisation. Figure \ref{fig:wavelets_spatial} shows an example of the spatial localisation of spin wavelets. Unlike scalar wavelets which are real-valued, spin wavelets are complex-valued. Figure \ref{fig:wavelets_harmonic} shows an example of the harmonic localisation of spin wavelets (for the wavelets used in this work). Figure \ref{fig:cmb_decomp} shows an example of spin-2 wavelet decomposition as applied to a simulated CMB polarisation field \(P\). By the construction of the spin-2 wavelets, the real and imaginary parts of the complex wavelet coefficient maps of \(P\) are respectively scalar wavelet transforms of \(E\) and \(B\) fields (with a different scalar wavelet). It can be seen that the spin-2 wavelet transform in the Spin-SILC method carries out the decomposition of the CMB polarisation into \(E\) and \(B\) modes. Details about the use of Spin-SILC on partial sky observations are given in \S~\ref{sec:cut_sky}.

\section{Method}
\label{sec:method}

We start by outlining the Spin-SILC algorithm. The steps are explained in more detail in the subsequent subsections (\S~\ref{sec:input} to \ref{sec:synthesis}). The use of Spin-SILC on partial sky observations is discussed in \S~\ref{sec:cut_sky}. We discuss our numerical implementation in \S~\ref{sec:num_imp}.
\begin{enumerate}
\renewcommand{\theenumi}{(\arabic{enumi})}
\item The raw input data are full-sky frequency maps of the anisotropies in the linear polarisation of the CMB, \ie Stokes \(Q\) and \(U\) fields. These maps use the HEALPix format \mbox{\citep{2005ApJ...622..759G}}. (See \S~\ref{sec:input}.) The model we employ for the raw data is explained in \S~\ref{sec:model}.
\item The maps are ``pre-processed'' by inpainting in a small point source mask and each convolved to have the same effective beam (see \S~\ref{sec:process}).
\item At each frequency band, the complex spin-2 polarisation field \(P = Q + iU\) is formed. Each \(P\) map is converted into a set of complex-valued spin-2 wavelet coefficient maps. This separates both the scale and orientation of structure within each map. These wavelet coefficient maps are sampled according to the sampling theorem of \citet{2015ISPL...22.2425M}. (See \S~\ref{sec:analysis}.)
\item A spin-2 ILC method is then applied separately to each wavelet scale and orientation. For each scale and orientation, the multifrequency wavelet coefficient maps are weighted and added to form a single (complex-valued) wavelet coefficient map that contains mainly CMB signal, as well as some residual foreground and noise. These weights are allowed to vary at each wavelet coefficient. The calculation of these weights is explained in \S~\ref{sec:ilc}.
\item By the construction of the spin-2 wavelets we use, the real and imaginary parts of the final ILC wavelet coefficient maps are respectively ILC estimates of the scalar wavelet transforms of the CMB \(E\) and \(B\) maps (with a different scalar wavelet). Therefore, the real and imaginary parts are separately synthesised with scalar wavelets to form the final products: full-sky maps of the CMB \(E\) and \(B\) anisotropies (with some residual foreground and noise). (See \S~\ref{sec:synthesis}.) \(Q\) and \(U\) maps are also formed by a standard spin-2 inverse wavelet transform of the ILC results (see \S~\ref{sec:spin_synthesis}). All final maps use the HEALPix format.
\end{enumerate}

\subsection{Input data}
\label{sec:input}

Our CMB polarisation map products use full-mission 2015 release {\Planck} \(Q\) and \(U\) polarisation maps as their input\footnote{\url{http://pla.esac.esa.int/pla}} \citep{2015arXiv150201585P,2015arXiv150201587P}. All seven polarisation frequency channels are used. At 70 GHz, we use the higher-resolution version at \(N_\mathrm{side} = 2048\). As noted in \citet{2015arXiv150201582P}, the 100, 143 and 217 GHz polarisation maps have been high-pass filtered due to insufficient characterisation of residual systematic effects on large scales, in particular leakage between temperature and polarisation measurements. We therefore follow \citet{2015arXiv150205956P} in also high-pass filtering the spherical harmonic coefficients of our output data products with a harmonic cosine filter:
\begin{equation}\label{eq:filter}
w_\ell =\begin{cases}
0, & \text{if \(\ell < 20\)}, \\
\frac{1}{2}\left[1 - \cos\left(\frac{\pi}{20}(\ell - 20)\right)\right], & \text{if \(20 \leq \ell \leq 40\)}, \\
1, & \text{otherwise}.
\end{cases}
\end{equation}

We use the full-mission Full Focal Plane 8 (FFP8) simulations \citep{2015arXiv150906348P} with lensed scalar perturbations and without bandpass mismatch. \kkr{These consist of a superposition of a CMB realisation, a noise realisation and full simulations of diffuse and point source astrophysical foregrounds.}

While we do not expect an algorithm developed for next generation precision CMB polarisation observations to demonstrate its full capabilities with the {\Planck} dataset, this setting comprises the best publicly-available simulations and data, and benefits from the availability of comparison data products from well-studied and highly tested component separation algorithms used by the Planck Collaboration. Thus we use the {\Planck} setting to benchmark our algorithm.

\subsection{Data model}
\label{sec:model}

Each of the full-sky Stokes \(Q\) and \(U\) polarisation maps (\(X = Q, U\)) can be independently\footnote{The independent modelling is based on the accurate assumption that any mixing of \(Q\) and \(U\) modes in their measurement has been previously corrected in any given experiment.} physically modelled \citep[\eg][]{2013MNRAS.435...18B} as
\begin{equation}\label{eq:model}
X^{\mathrm{OBS},c}(\hat{\vec{n}}) = \int_{\hat{\vec{n}}'} \mathrm{d}\hat{\vec{n}}' B^c(\hat{\vec{n}},\hat{\vec{n}}') X^{\mathrm{SIG},c}(\hat{\vec{n}}') + X^{\mathrm{N},c}(\hat{\vec{n}}),
\end{equation}
where the signal component can further be decomposed as
\begin{equation}\label{eq:signal}
X^{\mathrm{SIG},c}(\hat{\vec{n}}) = a^c X^{\mathrm{CMB}}(\hat{\vec{n}}) + X^{\mathrm{FG},c}(\hat{\vec{n}}).
\end{equation}
\(X^{\mathrm{CMB}}(\hat{\vec{n}})\) is the CMB component at a point on the sky \(\hat{\vec{n}}\). \(X^{\mathrm{FG},c}(\hat{\vec{n}})\) and \(X^{\mathrm{N},c}(\hat{\vec{n}})\) are respectively the foreground and detector noise components for frequency channel \(c\). \(a^c\) is the calibration coefficient for the CMB for each channel. The overall signal component is smoothed by a beam function \(B^c(\hat{\vec{n}},\hat{\vec{n}}')\) due to the finite resolution of the observations. However, the noise component is not smoothed by the beam. Here we assume the beam to be circularly symmetric. Therefore, the beam can be represented as a sum over Legendre polynomials,
\begin{equation}\label{eq:beam}
B^c(\hat{\vec{n}},\hat{\vec{n}}') = \sum_{\ell=0}^\infty \frac{2\ell+1}{4\pi} B_\ell^c P_\ell(\hat{\vec{n}}.\hat{\vec{n}}').
\end{equation}
We can recast Eq.~\eqref{eq:model} in the scalar spherical harmonic representation as
\begin{equation}\label{eq:model_alm}
\prescript{}{0}{a}_{\ell m}^{\mathrm{OBS},c} = a^c B_\ell^c \prescript{}{0}{a}_{\ell m}^\mathrm{CMB} + B_\ell^c \prescript{}{0}{a}_{\ell m}^{\mathrm{FG},c} + \prescript{}{0}{a}_{\ell m}^{\mathrm{N},c}
\end{equation}
where \(\prescript{}{0}{a}_{\ell m}\) are the coefficients of scalar spherical harmonics \(\prescript{}{0}{Y}_{\ell m}(\hat{\vec{n}})\).

The above is a useful representation for the data pre-processing in \S~\ref{sec:process}. However, the novelty of Spin-SILC is to develop a component separation algorithm that directly makes use of the spin properties of the CMB polarisation field. The Stokes \(Q\) and \(U\) parameters are defined with respect to a fixed coordinate system on the sky. They can be identified by their respective `+' and `\(\times\)' patterns, as seen in the top row of Fig.~\ref{fig:cmb_decomp}. However, there is no rotationally-invariant measure of the power as a function of scale. In order to address this, we can first form the complex-valued, spin \(\pm2\) polarisation field
\begin{equation}\label{eq:pol}
\begin{split}
\tensor[_{\pm2}]{P}{}(\hat{\vec{n}}) &= Q(\hat{\vec{n}}) \pm iU(\hat{\vec{n}}) \\
&= \sum_{\ell m} \tensor[_{\pm2}]{a}{_{\ell m}}\,\tensor[_{\pm2}]{Y}{_{\ell m}}(\hat{\vec{n}}).
\end{split}
\end{equation}
The spin \(\pm2\) property implies that \(\prescript{}{\pm2}{P}(\hat{\vec{n}})\) transforms under local rotations of angle \(\psi\) via \(\prescript{}{\pm2}{P} \rightarrow e^{-is\psi} \prescript{}{\pm2}{P}\), where spin number \(s = \pm2\). The second equality therefore follows by expanding the spin field \(\prescript{}{\pm2}{P}(\hat{\vec{n}})\) in the basis of spin spherical harmonics \(\prescript{}{\pm2}{Y}_{\ell m}(\hat{\vec{n}})\) with spin spherical harmonic coefficients \(\prescript{}{\pm2}{a}_{\ell m}\).

We can then define the scalar \(E\) and pseudo-scalar \(B\) fields in harmonic space as
\begin{equation}\label{eq:eb_fields}
\begin{split}
&\prescript{}{0}{E}_{\ell m} = -\frac{1}{2} (\prescript{}{2}{a}_{\ell m} + \prescript{}{-2}{a}_{\ell m}) \\
&\prescript{}{0}{B}_{\ell m} = \frac{i}{2} (\prescript{}{2}{a}_{\ell m} - \prescript{}{-2}{a}_{\ell m}).
\end{split}
\end{equation}
By construction, the \(E\) and \(B\) fields are real-valued and allow the calculation of rotationally-invariant angular power spectra. \(E\) modes are identified by the polarisation strength increasing in a direction parallel or perpendicular to the sense of the polarisation; it is the curl-free component of the spin-2 signal. \(B\) modes are identified by the polarisation strength increasing in a direction unaligned to the sense of the polarisation; it is the divergence-free component of the spin-2 signal. \(E\) and \(B\) modes therefore respectively separate the underlying field into parity-even and parity-odd components.

\subsection{Data pre-processing}
\label{sec:process}

The input frequency \(Q\) and \(U\) maps are diffusively inpainted in a small point source mask following the method employed by \citet{2015arXiv150201592P}. This recognises that the ILC fails when the CMB is obscured by bright extragalactic polarised sources. The inpainting removes these sources and replaces them with an extrapolation of the surrounding signal. The mask is the union of the {\Planck} LFI and HFI point source masks, which are constructed from the Second Planck Catalogue of Compact Sources (PCCS2) \citep{2015arXiv150702058P}\footnote{The details of their construction are given within the FITS files. They can be downloaded from \url{http://pla.esac.esa.int/pla}.}. It masks about 0.6\% of the whole sky, predominantly along the Galactic equator.

After the inpainting, we convert all the input frequency maps to the same resolution by performing a deconvolution/convolution procedure that gives scalar spherical harmonic coefficients
\begin{equation}\label{eq:conv}
\prescript{}{0}{a}_{\ell m}^c = \frac{B_\ell^\mathrm{EFF}}{B_\ell^c} \prescript{}{0}{a}_{\ell m}^{\mathrm{OBS},c} \, ,
\end{equation}
where \(B_\ell^\mathrm{EFF}\) is the beam transfer function giving the resolution at which we perform the ILC. For {\Planck} data, we use a Gaussian beam with a FWHM of 5\arcmin\ as our input beam. Our final map products are re-convolved to a {10\arcmin} beam in order to suppress residual noise. The beams we deconvolve \(B_\ell^c\) are taken from the Reduced Instrument Model (RIMO)\footnote{{\Planck} 2015 Release Explanatory Supplement: The 2015 instrument model (\url{http://wiki.cosmos.esa.int/planckpla2015/index.php/The_RIMO}).}. For the LFI beams, we use Gaussian approximations with FWHM {32.33\arcmin}, {27.01\arcmin} and {13.25\arcmin} for 30, 44 and 70 GHz respectively. Following \citet{2014A&A...571A..12P,2016arXiv160101322R}, the deconvolved beams are thresholded such that the \(B_\ell^c\) is set to the value given in the RIMO or 0.001, whichever is larger. This suppresses noise within the ILC method.

\subsection{Spin wavelet analysis}
\label{sec:analysis}

The spin wavelet ILC method requires the decomposition of each band-limited, complex-valued polarisation map \(\prescript{}{2}{P}^c(\hat{\vec{n}})\) (as formed by Eq.~\eqref{eq:pol}) into a set of spin wavelet coefficient maps \(W_P^{\prescript{}{2}{\Psi}^j}\). We use the spin, directional, scale-discretised wavelets of \citet{mcewen:s2let_spin, 2015arXiv150203120L, 2014IAUS..306...64M}. Following an introductory summary in \S~\ref{sec:spin_wav}, we now discuss some of the technical details of our wavelet implementation. We drop the \(c\) superscript on \(\prescript{}{2}{P}(\hat{\vec{n}})\) for the rest of this subsection since each frequency map is analysed using the same wavelets. We concentrate on the spin-2 wavelet transforms we use in Spin-SILC, but the wavelets we use can be generalised to arbitrary spin.

The spin wavelet coefficients are defined as the directional convolution of \(\prescript{}{2}{P}\) with spin wavelets \(\prescript{}{2}{\Psi}^j \) defined on the sphere \(\mathbb{S}^2\) (specifically those shown in Fig.~\ref{fig:wavelets_harmonic}), where index \(j\) denotes the wavelet scale. Like the scalar case \citep{2008MNRAS.388..770W, 2013SPIE.8858E..0IM}, spin, directional wavelets yield coefficients \(W_P^{\prescript{}{2}{\Psi}^j}(\hat{\vec{\rho}})\) that live on the space of three-dimensional rotations, \ie the rotation group SO(3):
\begin{equation}\label{eq:wav_coeffs}
W_P^{\prescript{}{2}{\Psi}^j}(\hat{\vec{\rho}}) \equiv \langle \prescript{}{2}{P}\,|\,\mathcal{R}_{\hat{\vec{\rho}}} \prescript{}{2}{\Psi}^j \rangle = \int_{\mathbb{S}^2} \mathrm{d}\hat{\vec{n}} \prescript{}{2}{P}(\hat{\vec{n}}) (\mathcal{R}_{\hat{\vec{\rho}}} \prescript{}{2}{\Psi}^j)^\ast(\hat{\vec{n}}),
\end{equation}
where \(\mathrm{d}\hat{\vec{n}}\) is the usual invariant measure on the sphere and $\cdot^\ast$ denotes complex conjugation. The rotation operator is defined by
\begin{equation}\label{eq:rot}
(\mathcal{R}_{\hat{\vec{\rho}}} \prescript{}{2}{\Psi}^j)(\hat{\vec{n}}) \equiv \prescript{}{2}{\Psi}^j (\mat{R}_{\hat{\vec{\rho}}}^{-1} \hat{\vec{n}}),
\end{equation}
where \(\mat{R}_{\hat{\vec{\rho}}}\) is the three-dimensional rotation matrix corresponding to \(\mathcal{R}_{\hat{\vec{\rho}}}\). In Eqs.~\eqref{eq:wav_coeffs} and \eqref{eq:rot}, \(\hat{\vec{\rho}} = (\theta,\phi,\chi) \in \mathrm{SO}(3)\) denotes the Euler angles (in the \(zyz\) convention) with colatitude \(\theta \in [0,\pi]\), longitude \(\phi \in [0,2\pi)\) and direction \(\chi \in [0,2\pi)\). In other words, the wavelet coefficients probe directional structure in \(\prescript{}{2}{P}\) with \(\chi\) corresponding to the orientation about each point \((\theta,\phi)\) on the sphere.

Spin, directional wavelets are defined by their spin spherical harmonic coefficients in factorised form:
\begin{equation}\label{eq:wavelets}
\tensor[{_2}]{\Psi}{_{\ell n}^j} \equiv \sqrt{\frac{2 \ell + 1}{8 \pi^2}} \kappa_\ell^j \tensor[{_2}]{\zeta}{_{\ell n}},
\end{equation}
where \(\kappa_\ell^j\) sets the harmonic localisation (Fig.~\ref{fig:wavelets_harmonic}) and \(\tensor[{_2}]{\zeta}{_{\ell n}}\) sets the directional localisation. Their spin properties are maintained by being built on the basis of spin spherical harmonics. Full details of their construction are given in \citet{mcewen:s2let_spin}. As in \citet{2016arXiv160101322R}, we flexibly control the harmonic localisation by using different values of the wavelet dilation parameter \(\lambda\) in different multipole regions. For the wavelets we use in Fig.~\ref{fig:wavelets_harmonic}, we use \(\lambda = 2, 1.3, 1.1\) with transitions at multipoles \(\ell = 512, 2015\). \kkr{The harmonic bounds of each wavelet for scale \(j\) are given by \((\ell_\mathrm{min}^j,\ell_\mathrm{max}^j) = (\lambda^{j'-1},\lambda^{j'+1})\), taking account of the different values of \(\lambda\) we use and the stitching-together of wavelets at \(\lambda\) transitions. The index \(j'\) refers to the original index of the wavelet scale as if only that single value of \(\lambda\) was used. Their peak response is at \(\lambda^{j'}\).} The details of our harmonic tiling are given in Table \ref{tab:wav_lims}. A single parameter \(N\) (at all scales) defines the number of directions into which each wavelet scale is localised.

We also use an axisymmetric scaling function \(\prescript{}{2}{\Phi}\) to form scaling coefficients \(W_P^{\prescript{}{2}{\Phi}}\) which characterise the largest-scale information (in this work for \(\ell < 64\)) and live on the sphere. This is motivated by testing in \citet{2016arXiv160101322R} that showed that the use of directionality on large scales in the ILC is not effective for CMB reconstruction. These wavelets (and the scaling function) satisfy the standard admissibility criterion for exact reconstruction, \ie no information is lost in the wavelet and inverse wavelet transforms of a band-limited spin signal. For the chosen band-limit, the smallest wavelet is harmonically-truncated. We choose not to use this wavelet, which means that exact reconstruction is only satisfied for \(\ell \leq 2048\). This allows the tapering of the smallest remaining wavelet (for \(2048 < \ell \leq 2253\)) to suppress the smallest-scale power in the algorithm.

In order to apply the ILC algorithm, the above continuous wavelet coefficients must be discretised. Since they live on the rotation group SO(3), we represent them using the sampling scheme of \citet{2015ISPL...22.2425M}, which is itself a generalisation of the sampling scheme of \citet{2011ITSP...59.5876M}. Since the wavelets are band-limited, we use a multi-resolution scheme where each wavelet scale \(j\) is pixellated with a minimal number of samples. This means that each wavelet coefficient map \(W_P^{\prescript{}{2}{\Psi}^j}\) (band-limited at \(\ell_\mathrm{max}^j\)) is only evaluated at samples \((\theta_t^j, \phi_p^j, \chi_n)\), where \(t \in \{0,1,\ldots,\ell_\mathrm{max}^j\}\), \(p \in \{0,1,\ldots,2\ell_\mathrm{max}^j\}\) and \(n \in \{0,1,\ldots,N-1\}\). In this way, each spin wavelet coefficient map can be separated into \(N\) spin, directional wavelet coefficient maps \(W_{jnk}^{\prescript{}{2}{P}}\) according to the value of \(n\), where \(k\) indexes pixel number according to samples \((\theta_t^j, \phi_p^j)\) on the sphere. It follows that each input frequency \(P\) map has been decomposed into wavelet coefficient maps, each localised according to harmonic scale \(j\) and orientation of structure \(n\), while maintaining spatial localisation (pixel number \(k\)). (See Fig.~\ref{fig:cmb_decomp} for a demonstration of this decomposition.)

\subsection{ILC method}
\label{sec:ilc}

Following the spin, directional wavelet analysis of the input \(P\) maps (see \S~\ref{sec:analysis}), there is a spin, directional wavelet coefficient map \(W_{jnk}^{\prescript{}{2}{P},c}\) for each channel \(c\), scale \(j\) and orientation \(n\) with a pixel index \(k\). With this compact notation, we conflate the scaling coefficient maps with the wavelet coefficient maps as the ILC method applies in exactly the same way. We develop the spin wavelet ILC method by an extension of the scalar wavelet ILC method we developed in \citet{2016arXiv160101322R} to operate on the complex-valued wavelet coefficient maps we now have. Similar to \citet{2016arXiv160101515F}, we consider the complex spin signal $\prescript{}{\pm2}{P}$ rather than considering scalar fields independently (\eg $Q$ and $U$ independently or $E$ and $B$ independently). However, unlike \citet{2016arXiv160101515F}, who work jointly on \(Q\) and \(U\) maps in real space, we work in wavelet space, where spatial, scale and directional localisation is possible. The most general extension of the scalar ILC is to estimate the CMB at each wavelet scale and orientation as a sum of wavelet coefficient maps for each frequency with complex-valued weights \(\omega_{jnk}^c\):
\begin{equation}\label{eq:ilc_sum}
W_{jnk}^{\prescript{}{2}{P},\mathrm{ILC}} \equiv \sum_{c=1}^{N_c} \omega_{jnk}^c W_{jnk}^{\prescript{}{2}{P},c}\,,
\end{equation}
where \(N_c\) is the number of input channels.

In order to recover an unbiased estimate of the CMB, we impose a constraint on the weights such that
\begin{equation}\label{eq:constraint}
\sum_{c=1}^{N_c} a^c \omega_{jnk}^c = 1 + 0i,
\end{equation}
where we remind the reader that \(a^c\) is the real-valued set of calibration coefficients for the CMB \(Q\) and \(U\) maps introduced in Eq.~\eqref{eq:signal}. In order to calculate the weights at each pixel \(k\), we choose to minimise the covariant quantity \(\left\langle {\big | W_{jnk}^{\prescript{}{2}{P},\mathrm{ILC}}\big |}^2 \right\rangle\) with respect to the complex-valued weights \(\omega_{jnk}^c\) themselves, under the constraint in Eq.~\eqref{eq:constraint}. This minimisation can be carried out with complex Lagrange multipliers (similarly to the scalar case) giving complex-valued weights
\begin{equation}\label{eq:weights}
\omega_{jnk}^c = \frac{\sum_{c'=1}^{N_c} (R_{jnk}^{-1})^{cc'} a^{c'}}{\sum_{c=1}^{N_c} \sum_{c'=1}^{N_c} a^c (R_{jnk}^{-1})^{cc'} a^{c'}},
\end{equation}
where the true covariance matrices at scale \(j\), orientation \(n\) and pixel \(k\), \((R_{jnk})^{cc'} = \left\langle {W_{jnk}^{\prescript{}{2}{P},c}}^* W_{jnk}^{\prescript{}{2}{P},c'} \right\rangle\) (where the angled brackets indicate an ensemble average, although in practice we empirically estimate these covariances as explained below). In this work, we assume \(a^c = 1,\,\forall\,c\), \ie that the CMB is perfectly calibrated in the data we use.

There are two main consequences from minimising the quantity we choose. \kkr{First, as in scalar SILC, we assume that the CMB and foregrounds and the CMB and noise are respectively uncorrelated. It follows that the ensemble cross-term between CMB and residual contamination is zero and, since the CMB is conserved by the constraint in Eq.~\eqref{eq:constraint}, we are minimising only the variance of the error in CMB reconstruction.} Second, by minimising the variance of the full complex-valued spin-2 wavelet coefficients using complex weights, we are in turn jointly minimising the variance of the \(E\) and \(B\) modes. This is thanks to the construction of the spin wavelets we use, as discussed in \S~\ref{sec:spin_synthesis}. Unlike a foreground cleaning algorithm acting on the \(Q\) and \(U\) or \(E\) and \(B\) maps separately, this approach ensures that all the information (\ie from the multiple polarisation channels and the \(Q\) and \(U\) cross terms) is used to jointly construct clean CMB polarisation $P$ and its \(E\) and \(B\) modes.

As in scalar SILC, we estimate the covariance matrices \((R_{jnk})^{cc'}\) empirically on the data. We achieve this by replacing the appropriate ensemble average with a weighted average of the surrounding pixels. Specifically, we smooth the maps of covariance matrix elements with a Gaussian kernel in harmonic space. The size of this kernel is proportional to the size of the wavelet used at each scale\footnote{\(\mathrm{FWHM}^j = 50 \sqrt{\frac{1200}{N_\mathrm{samp}^j}}\). This value is the same as used in the NILC implementation on {\Planck} data.}. Full details of this empirical estimation of covariances and possible optimisations to the method are given in \citet{2016arXiv160101322R}.

\subsection{Spin wavelet synthesis to Stokes \(Q\) and \(U\) modes}
\label{sec:spin_synthesis}

Although a novelty of Spin-SILC is to simultaneously perform \(E\)-\(B\) decomposition and component separation (\ie to synthesise the ILC results directly to \(E\) and \(B\) maps as explained in \S~\ref{sec:synthesis}), one can also form \(Q\) and \(U\) maps. This is carried out by a single \mbox{spin-2} inverse wavelet transform. After the ILC method (see \S~\ref{sec:ilc}) has been applied to the frequency wavelet coefficient maps, there is one ILC estimate of the CMB \(P\) field (with some residual foreground and noise) at each wavelet scale and orientation \(W_{jnk}^{\prescript{}{2}{P},\mathrm{ILC}}\). Multiple orientations \(\chi_0,\chi_1,\ldots,\chi_{N-1}\) are combined at each scale to form wavelet coefficient maps \(W_P^{\prescript{}{2}{\Psi}^j,\mathrm{ILC}}(\hat{\vec{\rho}})\) that live on SO(3). We also have the scaling coefficient map \(W_P^{\prescript{}{2}{\Phi},\mathrm{ILC}}(\hat{\vec{n}})\) that lives on the sphere (and characterises the largest scales). In order to calculate our real space estimate of the CMB polarisation spin field \(\prescript{}{2}{P}^\mathrm{ILC}(\hat{\vec{n}})\) (and hence the Stokes parameters \(Q^\mathrm{ILC}(\hat{\vec{n}})\) and \(U^\mathrm{ILC}(\hat{\vec{n}})\)), we perform the following spin-2 inverse wavelet transform:
\begin{equation}\label{eq:spin_syn}
\begin{split}
\prescript{}{2}{P}^\mathrm{ILC}(\hat{\vec{n}}) =\,&Q^\mathrm{ILC}(\hat{\vec{n}}) + iU^\mathrm{ILC}(\hat{\vec{n}}) \\
= &\int_{\mathbb{S}^2} \mathrm{d}\hat{\vec{n}}' W_P^{\prescript{}{2}{\Phi},\mathrm{ILC}}(\hat{\vec{n}}') (\mathcal{R}_{\hat{\vec{n}}'} \prescript{}{2}{\Phi})(\hat{\vec{n}}) \\
&+ \sum_j \int_{\mathrm{SO}(3)} \mathrm{d}\hat{\vec{\rho}} W_P^{\prescript{}{2}{\Psi}^j,\mathrm{ILC}}(\hat{\vec{\rho}}) (\mathcal{R}_{\hat{\vec{\rho}}} \prescript{}{2}{\Psi}^j) (\hat{\vec{n}}),
\end{split}
\end{equation}
where $\mathrm{d}\hat{\vec{\rho}}$ is the usual invariant measure on the rotation group. We have used the same spin wavelets as in the wavelet analysis in \S~\ref{sec:analysis}. The final ILC \(Q\) and \(U\) maps are pixellated in the HEALPix format.

\subsection{Scalar wavelet synthesis to \(E\) and \(B\) modes}
\label{sec:synthesis}

A considerable advantage of the Spin-SILC method is that it simultaneously removes foreground and noise contamination from the cosmological signal and carries out the \(E\)-\(B\) decomposition discussed in \S~\ref{sec:model}. It achieves the latter by using a property of the spin scale-discretised wavelets that relates \(W_P^{\prescript{}{2}{\Psi}^j}(\hat{\vec{\rho}})\), the spin-2 wavelet transform of \(P\) to \(W_{\tilde{E}}^{\prescript{}{0}{\Psi}^j}(\hat{\vec{\rho}})\) and \(W_{\tilde{B}}^{\prescript{}{0}{\Psi}^j}(\hat{\vec{\rho}})\), the scalar wavelet transforms of \(\tilde{E}\) and \(\tilde{B}\):
\begin{equation}\label{eq:eb_sep}
\begin{split}
W_{\tilde{E}}^{\prescript{}{0}{\Psi}^j}(\hat{\vec{\rho}}) &= - \mathrm{Re}[W_P^{\prescript{}{2}{\Psi}^j}(\hat{\vec{\rho}})] \\
W_{\tilde{B}}^{\prescript{}{0}{\Psi}^j}(\hat{\vec{\rho}}) &= - \mathrm{Im}[W_P^{\prescript{}{2}{\Psi}^j}(\hat{\vec{\rho}})].
\end{split}
\end{equation}
The intermediate fields \(\tilde{E}\) and \(\tilde{B}\) are respectively related to \(E\) and \(B\) by a harmonic normalisation of their scalar spherical harmonic coefficients:
\begin{equation}\label{eq:intermediate}
\begin{split}
\tensor[_0]{E}{_{\ell m}} = \frac{1}{N_{\ell,2}} \tensor[_0]{\tilde{E}}{_{\ell m}} \\
\tensor[_0]{B}{_{\ell m}} = \frac{1}{N_{\ell,2}} \tensor[_0]{\tilde{B}}{_{\ell m}},
\end{split}
\end{equation}
where \(N_{\ell,s} = \sqrt{\frac{(\ell + s)!}{(\ell - s)!}}\).

The straightforward \(E\)-\(B\) decomposition is achieved by the construction of the wavelets and is discussed in detail in \citet{mcewen:s2let_spin} and \citet{ebsep}. In Eq.~\eqref{eq:eb_sep} the scalar wavelets \(\prescript{}{0}{\Psi}^j\) are spin-lowered versions of the spin-2 wavelets \(\prescript{}{2}{\Psi}^j\):
\begin{equation}\label{eq:spin_lower}
\prescript{}{0}{\Psi}^j (\hat{\vec{n}}) = {\overline{\eth}}^2 \prescript{}{2}{\Psi}^j (\hat{\vec{n}}).
\end{equation}
(An equivalent equation links the scalar and spin scaling functions.) \(\overline{\eth}\) is a first-order differential operator known as the spin-lowering operator since it lowers the spin of spherical harmonic functions: \(\overline{\eth} \prescript{}{s}{Y}_{\ell m}(\hat{\vec{n}}) = \frac{1}{N_{\ell,s}} \prescript{}{s-1}{Y}_{\ell m}(\hat{\vec{n}})\).

By applying Eq.~\eqref{eq:eb_sep}, we can separate the spin wavelet coefficient maps \(W_P^{\prescript{}{2}{\Psi}^j,\mathrm{ILC}}(\hat{\vec{\rho}})\) (defined in \S~\ref{sec:spin_synthesis}) into scalar wavelet coefficient maps of the intermediate \(\tilde{E}\) and \(\tilde{B}\) modes \(W_Y^{\prescript{}{0}{\Psi}^j,\mathrm{ILC}}(\hat{\vec{\rho}})\), where \(Y = \tilde{E},\tilde{B}\). An equivalent separation forms the scaling coefficient maps \(W_Y^{\prescript{}{0}{\Phi},\mathrm{ILC}}(\hat{\vec{n}})\). In order to calculate our real space estimates of the CMB \(\tilde{E}\) and \(\tilde{B}\) modes \(\prescript{}{0}{Y}^\mathrm{ILC}(\hat{\vec{n}})\), we perform inverse scalar wavelet transforms with the (spin-lowered) scalar scaling function \(\prescript{}{0}{\Phi}\) and scalar wavelets \(\prescript{}{0}{\Psi}^j\) (as defined in Eq.~\eqref{eq:spin_lower}):
\begin{equation}\label{eq:wav_syn}
\begin{split}
\prescript{}{0}{Y}^\mathrm{ILC}(\hat{\vec{n}}) = &\int_{\mathbb{S}^2} \mathrm{d}\hat{\vec{n}}' W_Y^{\prescript{}{0}{\Phi},\mathrm{ILC}}(\hat{\vec{n}}') (\mathcal{R}_{\hat{\vec{n}}'} \prescript{}{0}{\Phi})(\hat{\vec{n}}) \\
&+ \sum_j \int_{\mathrm{SO}(3)} \mathrm{d}\hat{\vec{\rho}} W_Y^{\prescript{}{0}{\Psi}^j,\mathrm{ILC}}(\hat{\vec{\rho}}) (\mathcal{R}_{\hat{\vec{\rho}}} \prescript{}{0}{\Psi}^j) (\hat{\vec{n}}).
\end{split}
\end{equation}
The output scalar spherical harmonic coefficients can be renormalised to the usual \(E\) and \(B\) fields by applying Eq.~\eqref{eq:intermediate}. The final ILC \(E\) and \(B\) maps are pixellated in the HEALPix format.

\subsection{Spin-SILC on partial sky observations}
\label{sec:cut_sky}

We have outlined above the Spin-SILC method specifically as it applies on the full sky. However, Spin-SILC is being primarily developed for application to future CMB polarisation experiments, which will have greater signal-to-noise and/or resolution, but will typically observe only part of the sky. The decomposition of Stokes \(Q\) and \(U\) measurements into \(E\) and \(B\) modes is essential for cosmological analyses, in particular for a measurement of the \(BB\) angular power spectrum. This is strictly well-defined only on the whole sky, as in Eq.~\eqref{eq:eb_fields}.

This decomposition is not well-defined if the input measurements only cover a part of the sky. However, following \citet{2003PhRvD..67b3501B}, a polarisation field on the cut-sky can be decomposed into a complete orthonormal basis defined by ``pure \(E\)'', ``pure \(B\)'' and ``ambiguous'' modes. Pure \(E\) modes have vanishing curl and are orthogonal to all \(B\) modes on the partial sky. Pure \(B\) modes have vanishing divergence and are orthogonal to all \(E\) modes on the partial sky. Ambiguous modes are all other modes, which will have both vanishing divergence and curl. It is the inability to distinguish ambiguous modes which leads to the problem of \(E\)-\(B\) leakage where ambiguous modes are erroneously counted as \(E\) or \(B\). However, following, \eg \citet{2003PhRvD..67b3501B,2007PhRvD..76d3001S}, if the pure \(B\) modes can be isolated, they will form an estimate of the cosmological \(B\) power, unbiased by \(E\)-\(B\) leaking.

\begin{figure}
\begin{tabular}{c}
\includegraphics[width=\columnwidth]{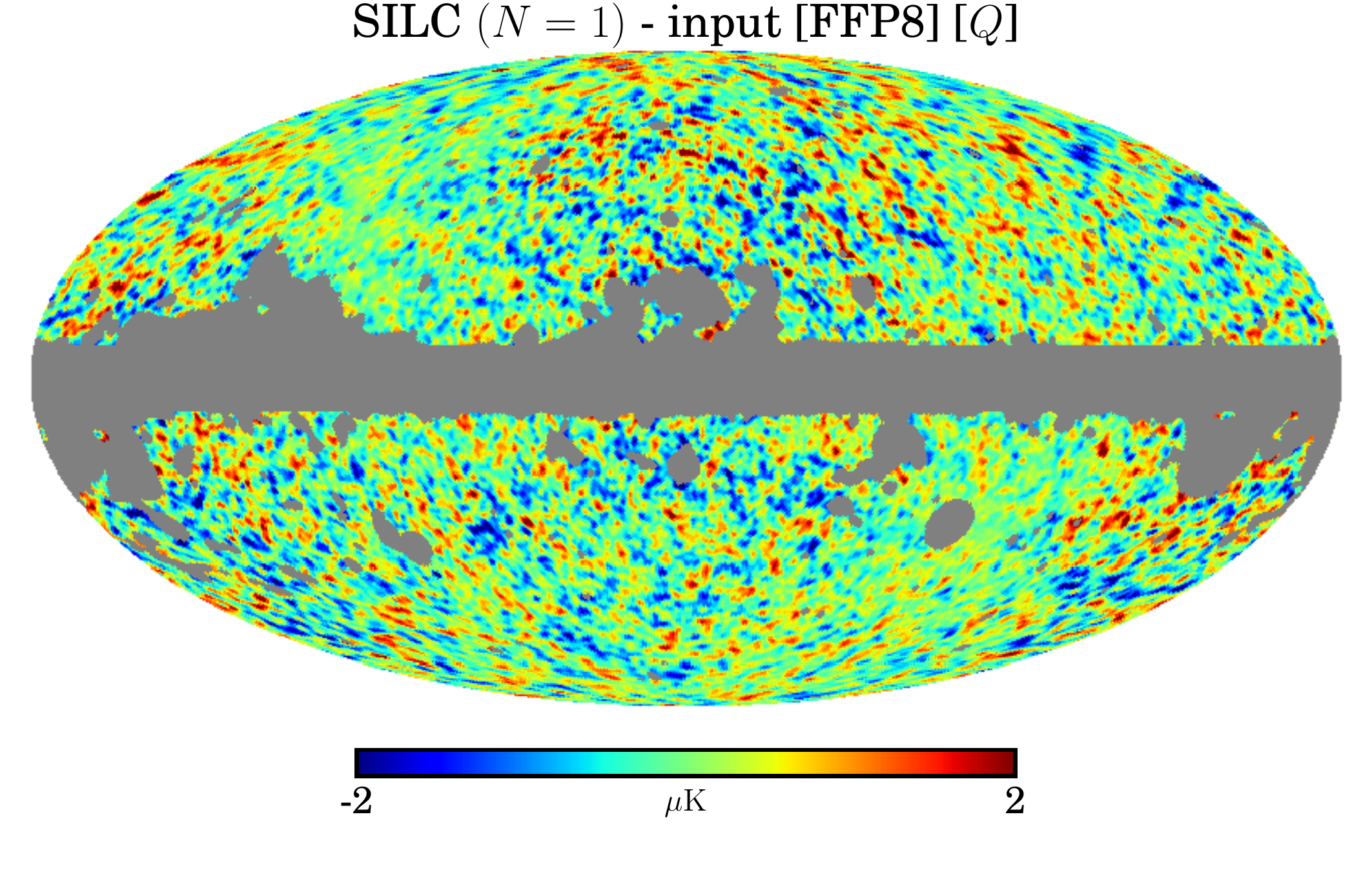} \\
\includegraphics[width=\columnwidth]{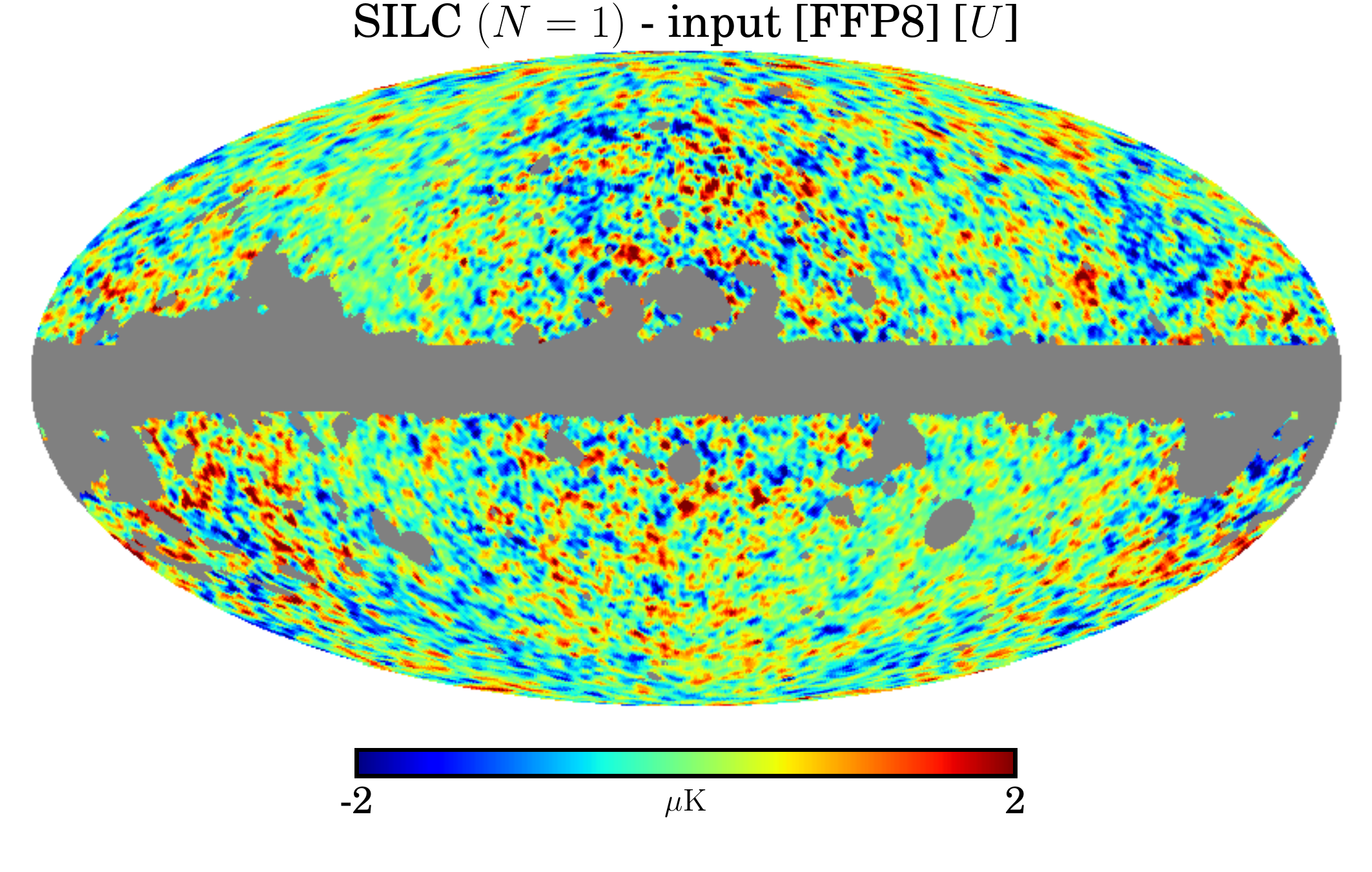}
\end{tabular}
\caption{Planck \textit{simulations.} Differences between output ILC and input CMB maps from FFP8 simulations with lensed scalar perturbations. The maps have been smoothed to \(\mathrm{FWHM} = 80\arcmin\) and downgraded to \(N_\mathrm{side} = 128\). The grey pixels are the UPB77 confidence mask from \citet{2015arXiv150205956P}, which masks the Galactic region in FFP8 simulations where foreground emission is strongest. \textit{From top to bottom,} we show differences in (a) Stokes \(Q\) and (b) Stokes \(U\) maps.}
\label{fig:ffp8N1diffMap}
\end{figure}

\citet{ebsep} have shown how the spin wavelets we use can be employed to construct estimates of the pure modes defined above on a masked sky. This builds on the work of, \eg \citet{2002PhRvD..65b3505L,2007PhRvD..76d3001S,2012PhRvD..86g6005G}. In particular, the spin wavelet pure-mode estimation of \citet{ebsep} requires only two additional wavelet transforms of the input \(Q\) and \(U\) data and a suitably apodised mask. The Spin-SILC method can be applied on partial sky observations by coherently combining the full-sky method with the pure mode estimation. The application of Spin-SILC to partial sky observations will be investigated in future work, providing the first integrated pipeline to simultaneously carry out \(E\)-\(B\) decomposition and foreground component separation for future CMB polarisation experiments.

\subsection{Numerical implementation}
\label{sec:num_imp}

\begin{figure}
\includegraphics[width=\columnwidth]{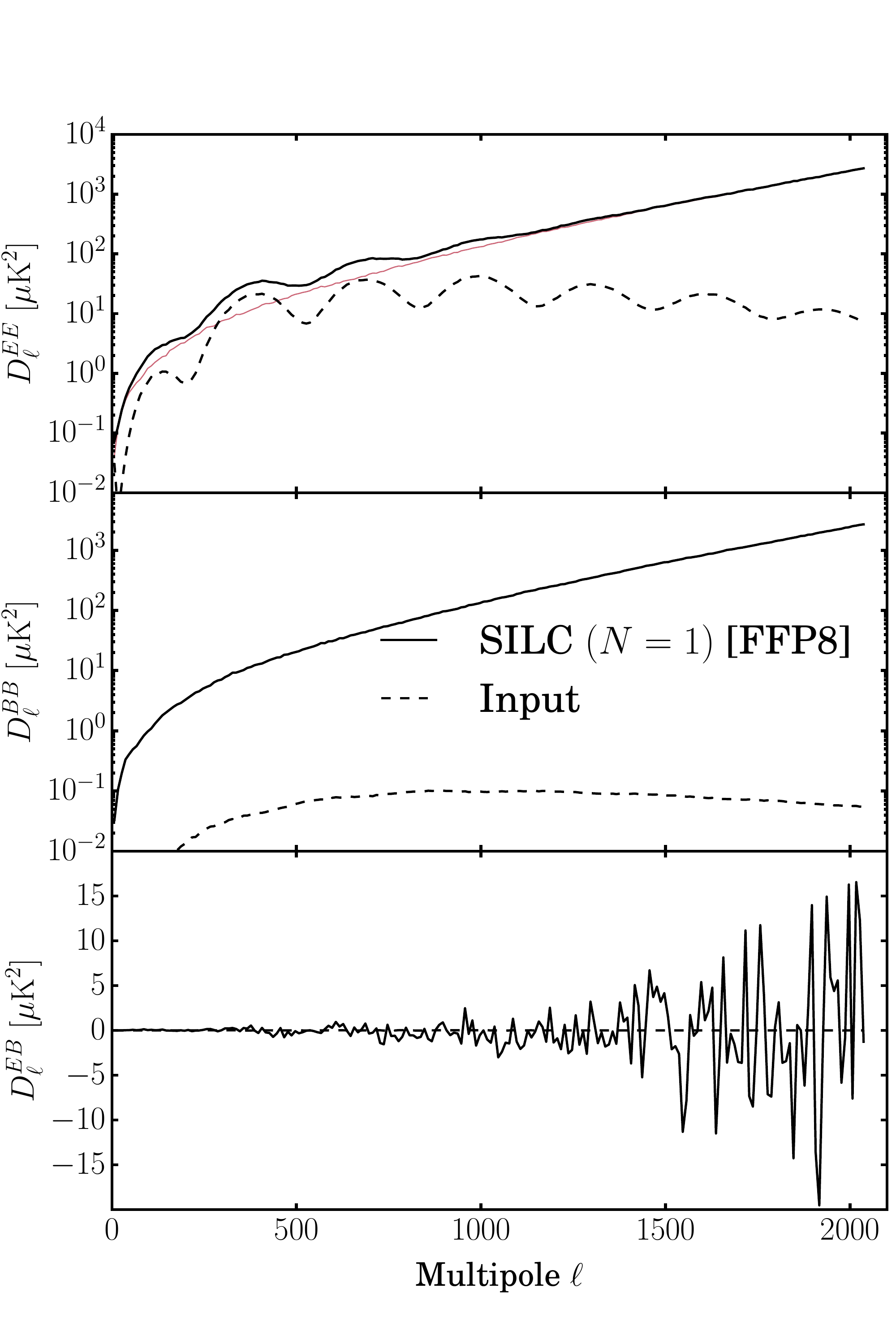}
\caption{Planck \textit{simulations.} \textit{From top to bottom,} (a) \(EE\), (b) \(BB\) and (c) \(EB\) angular power spectra comparing output ILC in the axisymmetric limit \((N = 1)\) to input CMB from FFP8 simulations with lensed scalar perturbations. \textit{In the top panel} (a), the thin red line shows residuals after subtracting the input CMB spectrum.}
\label{fig:ffp8N1spec}
\end{figure}

Spin-SILC is implemented in \texttt{Python} and is parallelised. At full {\Planck} resolution (\(N_\mathrm{side} = 2048\), \(\ell_\mathrm{max} = 2253\)), when run on a 60-core symmetric multiprocessor (SMP) with 1.5 TB RAM and a 24-core cluster node with 256 GB RAM\footnote{The exact specification for our infrastructure is an Intel Xeon E7-4890 2.8 GHz SMP with 4 \(\times\) 15-core CPUs with 25.6 GB RAM per core, and an Intel Xeon E5-2697 2.7 GHz node with 2 \(\times\) 12-core CPUs with 10.7 GB RAM per core.}, the pipeline takes approximately 1.5 hours per direction. The wavelet transforms in Spin-SILC are carried out using the latest version of the \texttt{S2LET}\footnote{\url{http://www.s2let.org}} code \citep{2013A&A...558A.128L,2015ISPL...22.2425M}, written in \texttt{C} with \texttt{Python} wrappers. This employs \texttt{SSHT}\footnote{\url{http://www.spinsht.org}} \citep{2011ITSP...59.5876M} and \texttt{SO3}\footnote{\url{http://www.sothree.org}} \citep{2015ISPL...22.2425M} to compute spin spherical harmonics and Wigner transforms exactly and efficiently. Spin-SILC is developed from the scalar SILC\footnote{\url{http://www.silc-cmb.org}} code \citep{2016arXiv160101322R} (which performs component separation on the temperature anisotropies of the CMB).

\section{Application to \textit{Planck} simulations}
\label{sec:simulations}

\begin{figure}
\includegraphics[width=\columnwidth]{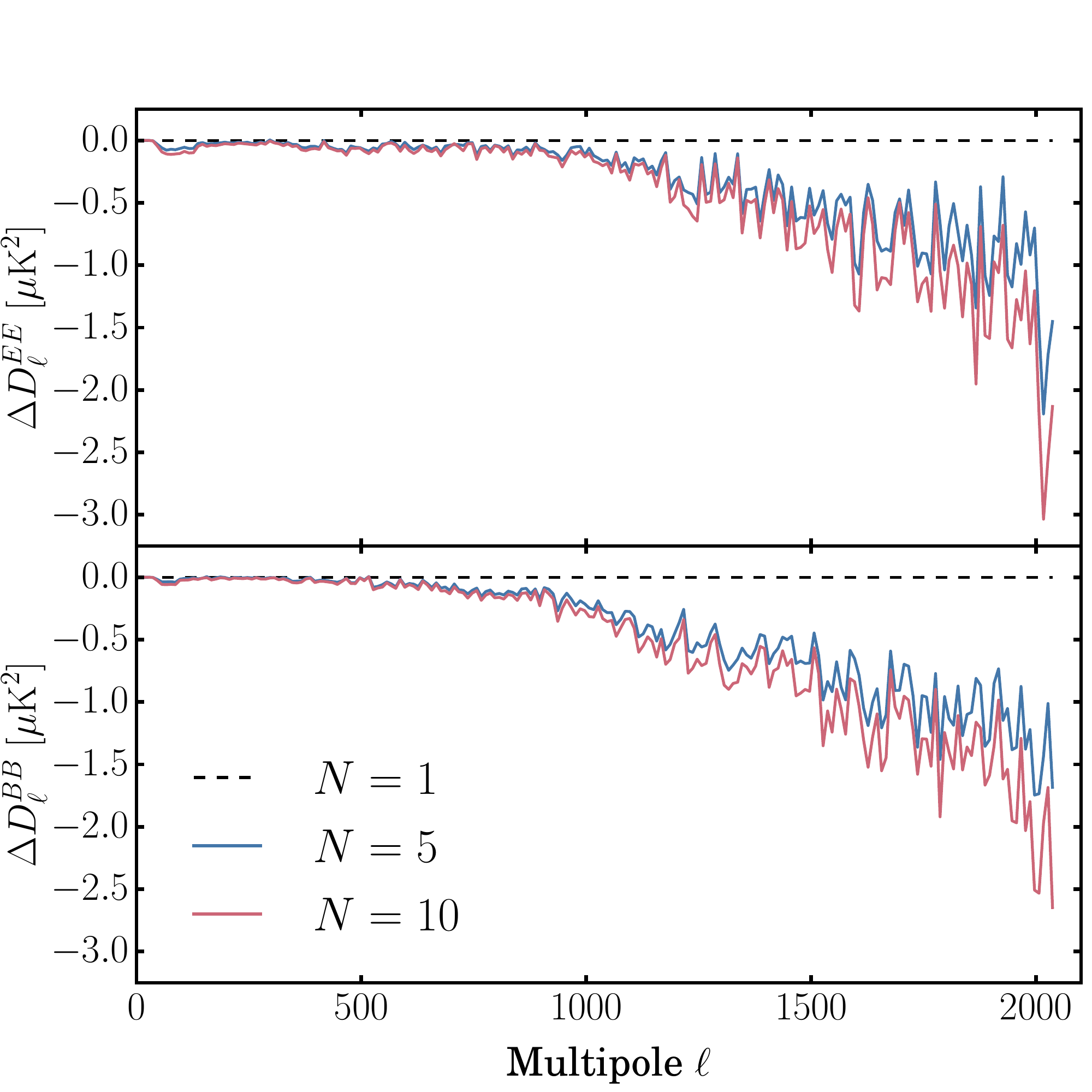}
\caption{Planck \textit{simulations.} Differences between angular power spectra of different values of \(N\) minus the axisymmetric limit \((N = 1)\). The input data are FFP8 simulations with lensed scalar perturbations. \textit{From top to bottom,} we show differences in (a) \(EE\) and (b) \(BB\) spectra. We note the small amplitude of the reductions in reconstruction residuals from increasing \(N\).}
\label{fig:ffp8N10spec}
\end{figure}

We tested Spin-SILC on the fiducial full-mission {\Planck} FFP8 simulated Stokes \(Q\) and \(U\) sky maps. We use simulations with lensed scalar perturbations. Figure~\ref{fig:ffp8N1diffMap} shows the differences between the reconstructed CMB (using \(N = 1\)) and the input simulated CMB. The two panels show the differences in \(Q\) and \(U\) maps, as this most directly compares to the input data. The most striking features are the reductions in residuals in the top left and bottom right corners, aligning with the Ecliptic poles. This reflects reduced noise residuals because there is less noise in the input data due to the scanning strategy of the {\Planck} satellite, which integrated for longer in those directions. In general, the difference maps are consistent with noise residuals. This is a consistent attribute of the {\Planck} polarisation datasets. Figure~\ref{fig:ffp8N1spec} compares the full-sky angular power spectra (\(D_\ell = \ell(\ell+1)C_\ell/2\pi\)) of the same reconstructed CMB and the input signal. The three panels respectively compare the \(EE\), \(BB\) and \(EB\) spectra, as these are the cosmologically-interesting observables. The \(EE\) and \(BB\) spectra are consistent with significant residual noise power due to the noisiness of the input maps, although the first four acoustic peaks of the \(EE\) spectrum are discernible nonetheless. The reconstructed \(EB\) spectrum is consistent with the zero input value.

We tested the impact of using directional spin wavelets in the Spin-SILC method with the simulated dataset. Figure~\ref{fig:ffp8N10spec} compares the differences in full-sky power spectra between the directional case (for \(N = 5,10\)) minus the axisymmetric limit (\(N = 1\)). The two panels compare \(EE\) and \(BB\) spectra. It can be seen that using more directional wavelets per wavelet scale reduces power spectrum reconstruction residuals with respect to the axisymmetric limit, very modestly on large scales and more so on small scales. However, the magnitude of these reductions is very small compared to the total power in the output ILC maps, which are dominated by residual instrumental noise; in the \(BB\) spectrum, the reduction is comparable to the magnitude of the input lensing signal. These results are fully expected following \citet{2016arXiv160101322R}, where it was found that the gains in component separation efficacy from employing directionality was marginal in the low signal-to-noise (S/N) regime.

\section{Application to \textit{Planck} data}
\label{sec:data}

\begin{figure}
\begin{tabular}{cc}
\includegraphics[width=0.5\columnwidth]{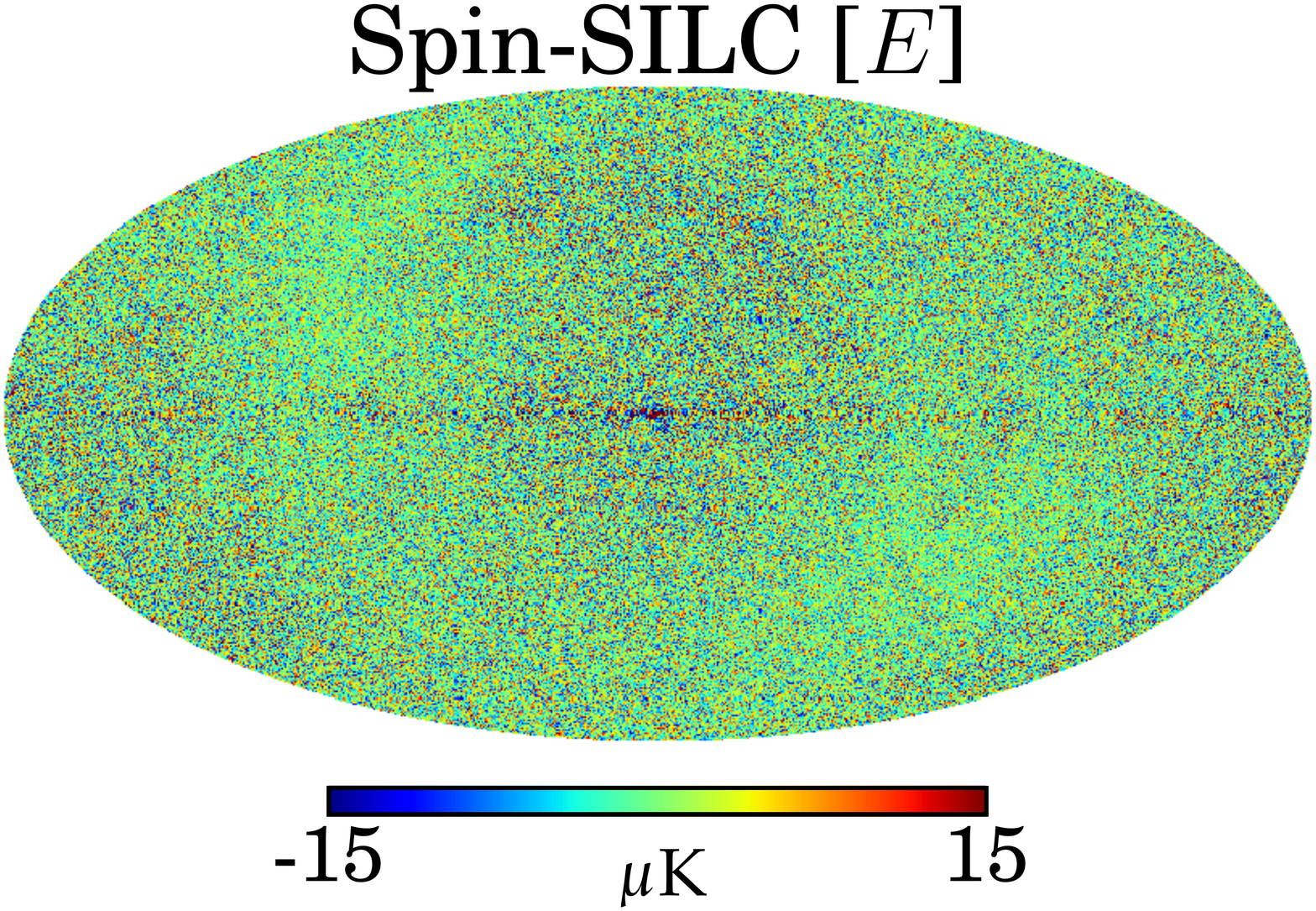} & \includegraphics[width=0.5\columnwidth]{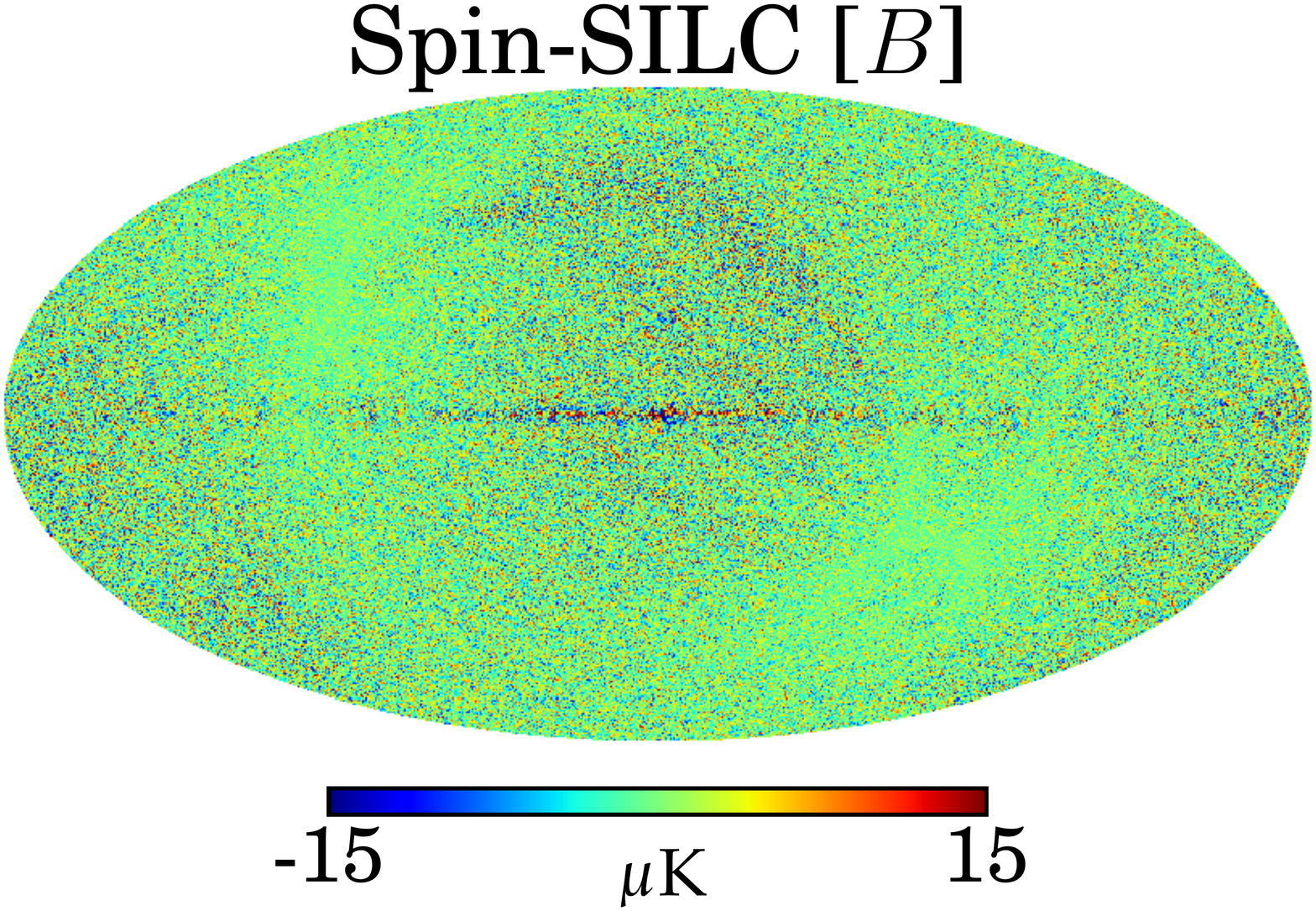}
\end{tabular}
\caption{Planck \textit{data.} \textit{From left to right,} (a) the CMB polarisation \(E\) map and (b) the CMB polarisation \(B\) map reconstructed using Spin-SILC in the axisymmetric limit \((N = 1, \mathrm{FWHM} = 10\arcmin, N_\mathrm{side} = 1024)\).}
\label{fig:EBmap}
\end{figure}

After testing Spin-SILC on the simulated dataset, we apply as input data the real full-mission {\Planck} Stokes \(Q\) and \(U\) maps. Figure~\ref{fig:EBmap} shows our main output data products: full-sky ILC estimates of the CMB polarisation \(E\) and \(B\) modes using Spin-SILC in the axisymmetric limit (\(N = 1\)). We show \(E\) and \(B\) maps in order to highlight the \(E\)-\(B\) decomposition from input \(Q\) and \(U\) maps that Spin-SILC automatically carries out thanks to the construction of the spin wavelets that we use (see \S~\ref{sec:analysis}). We reiterate that these maps have been high-pass filtered (for \(\ell < 40\)) in order to mitigate for residual systematics in the {\Planck} polarisation data (see \S~\ref{sec:input}). The maps are consistent with large levels of residual instrumental noise, with the scanning pattern of the {\Planck} satellite clearly visible. We also note the poor reconstruction in the Galactic plane, particularly towards the Galactic centre, where foreground emission is strongest and most complex.

\section{Comparison to previous work}
\label{sec:comp}

\begin{figure*}
\begin{tabular}{cc}
\includegraphics[width=\columnwidth]{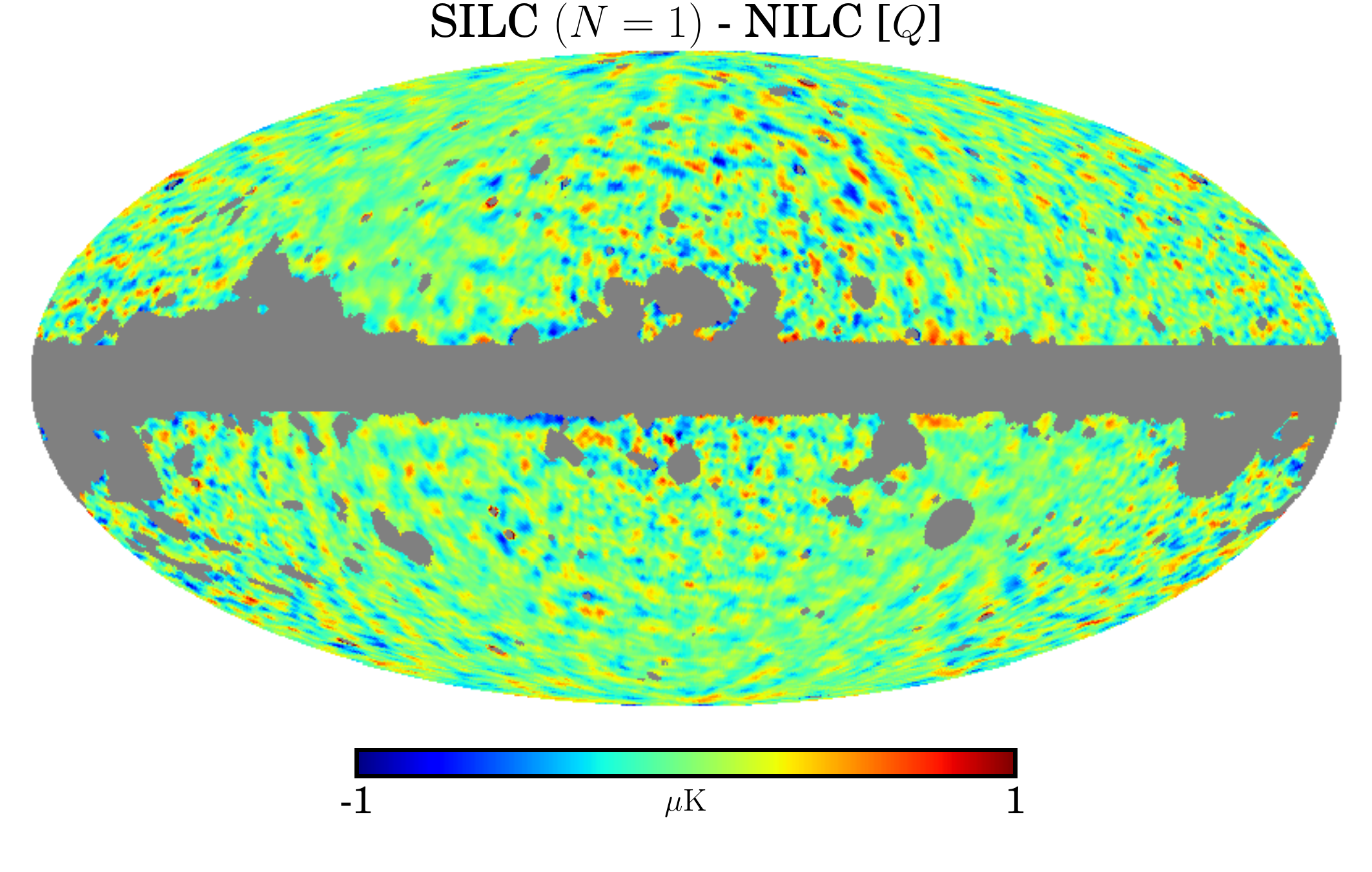} & \includegraphics[width=\columnwidth]{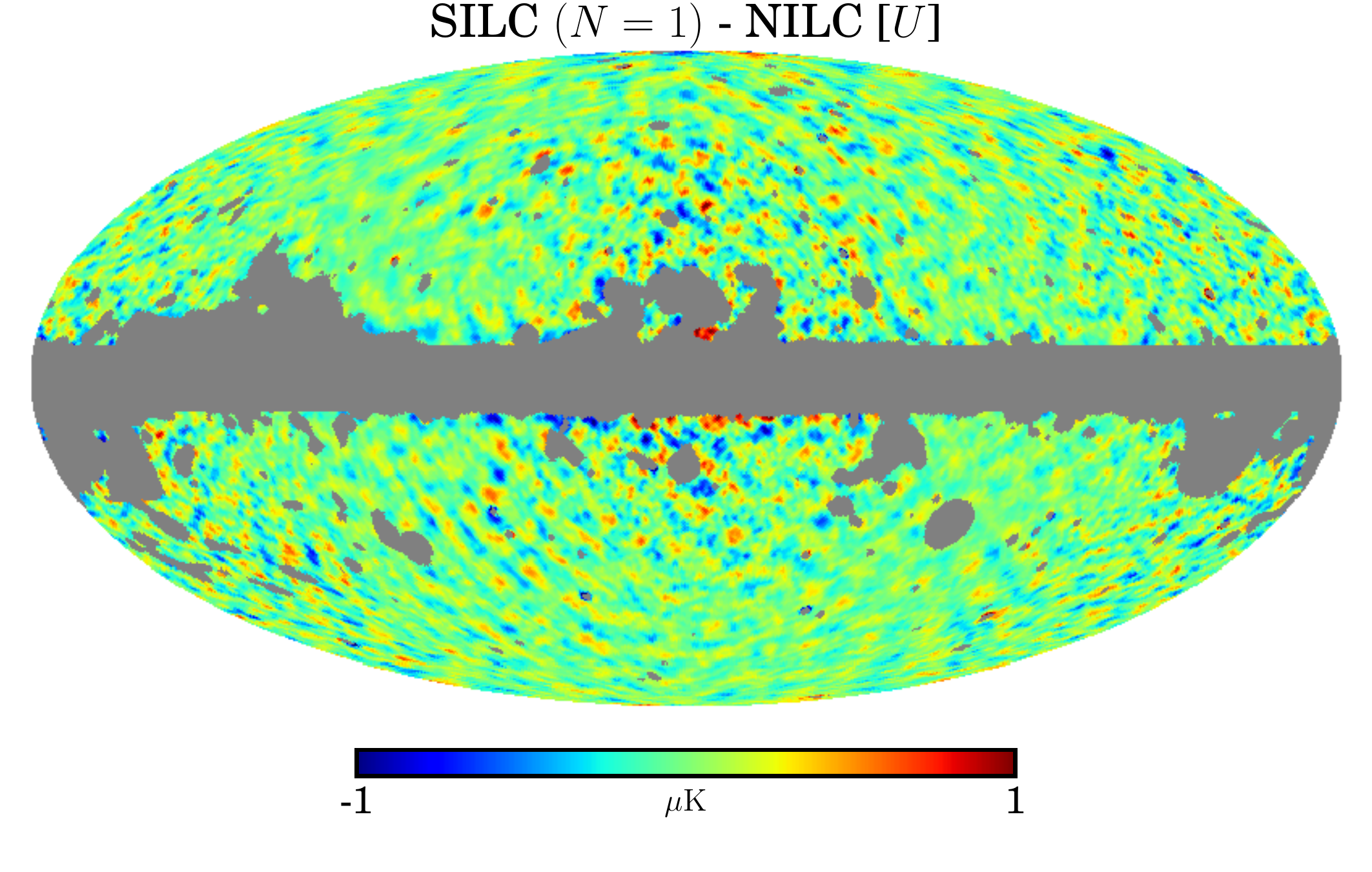} \\
\includegraphics[width=\columnwidth]{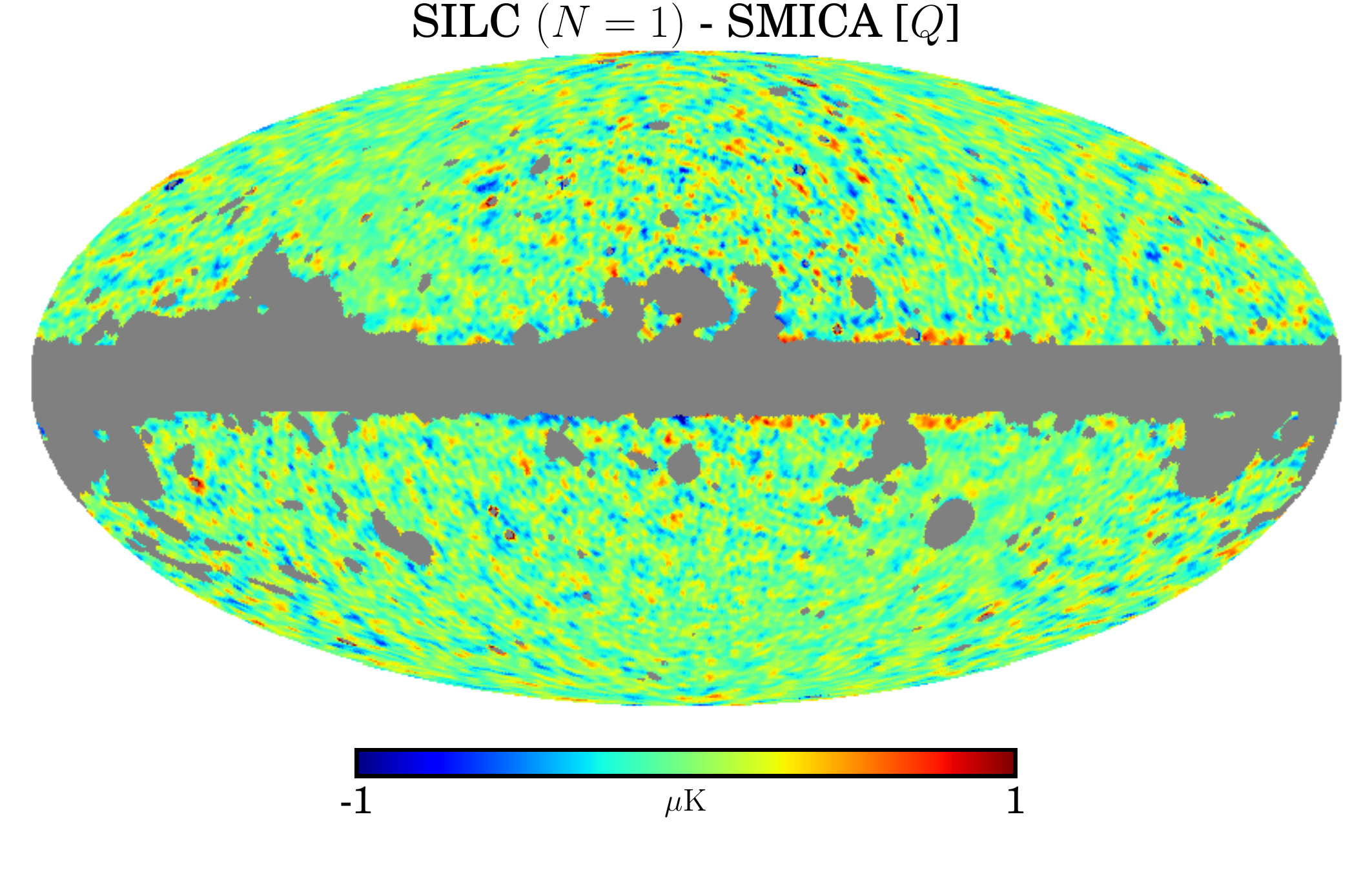} & \includegraphics[width=\columnwidth]{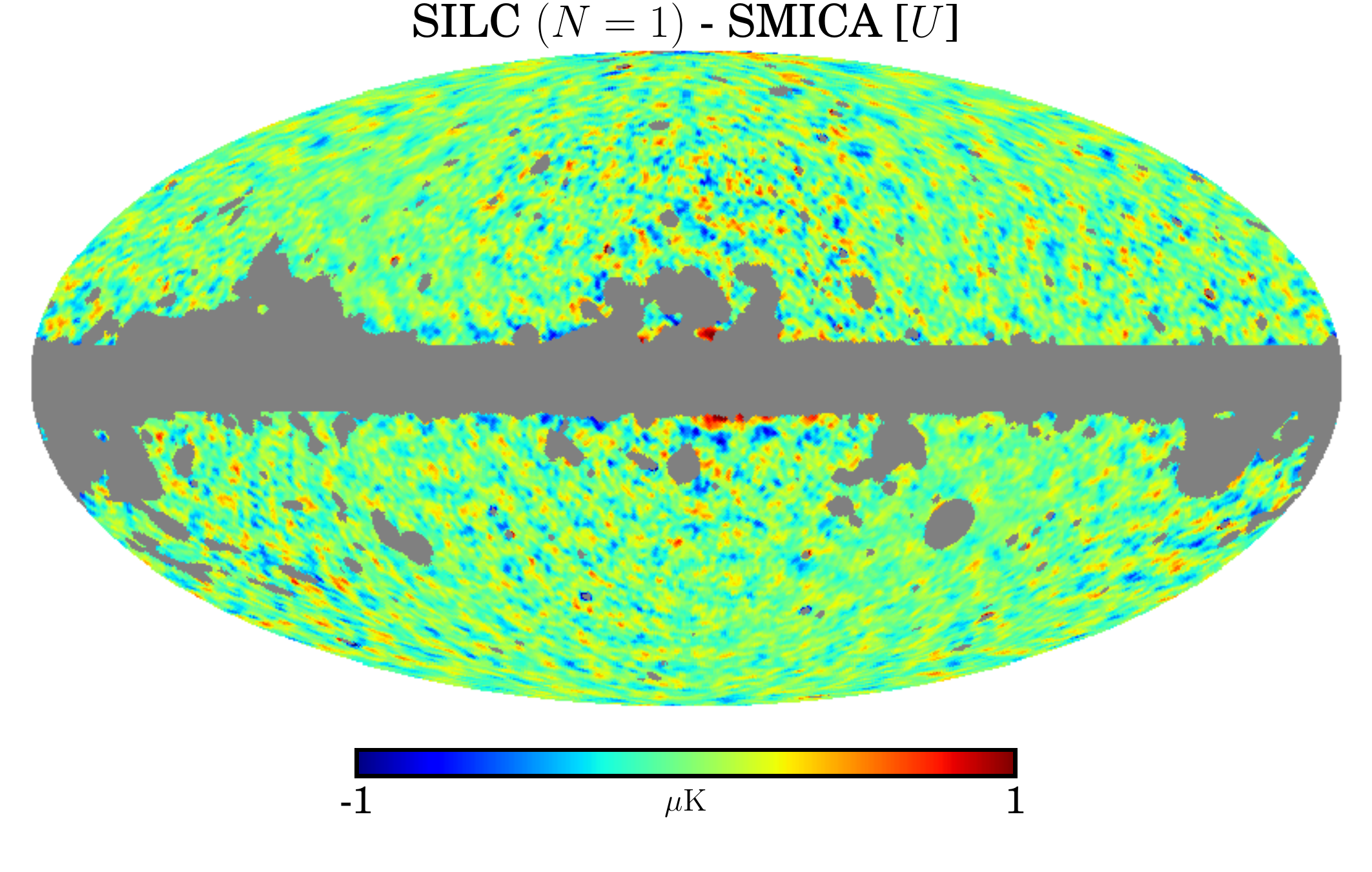} \\
\includegraphics[width=\columnwidth]{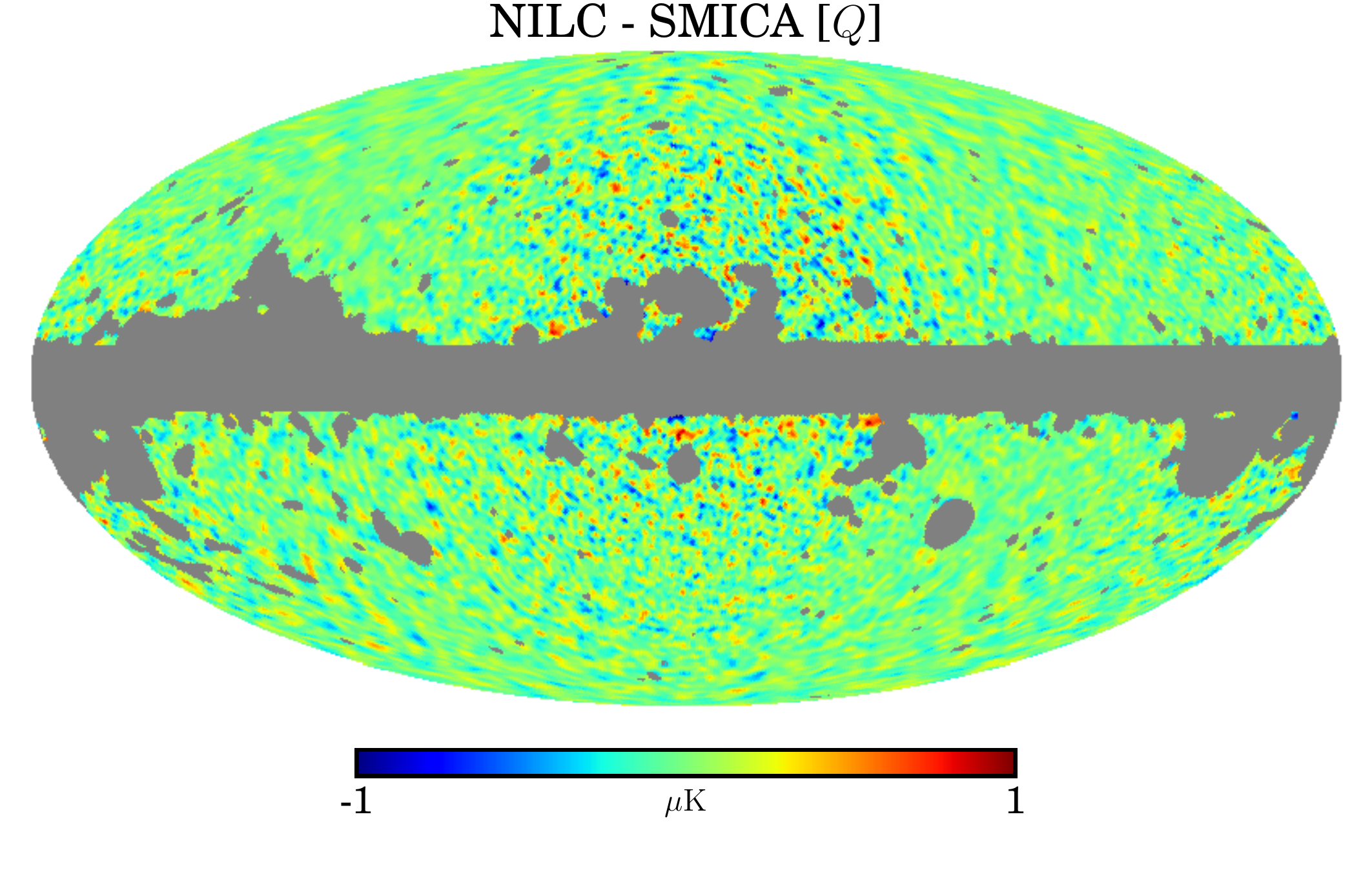} & \includegraphics[width=\columnwidth]{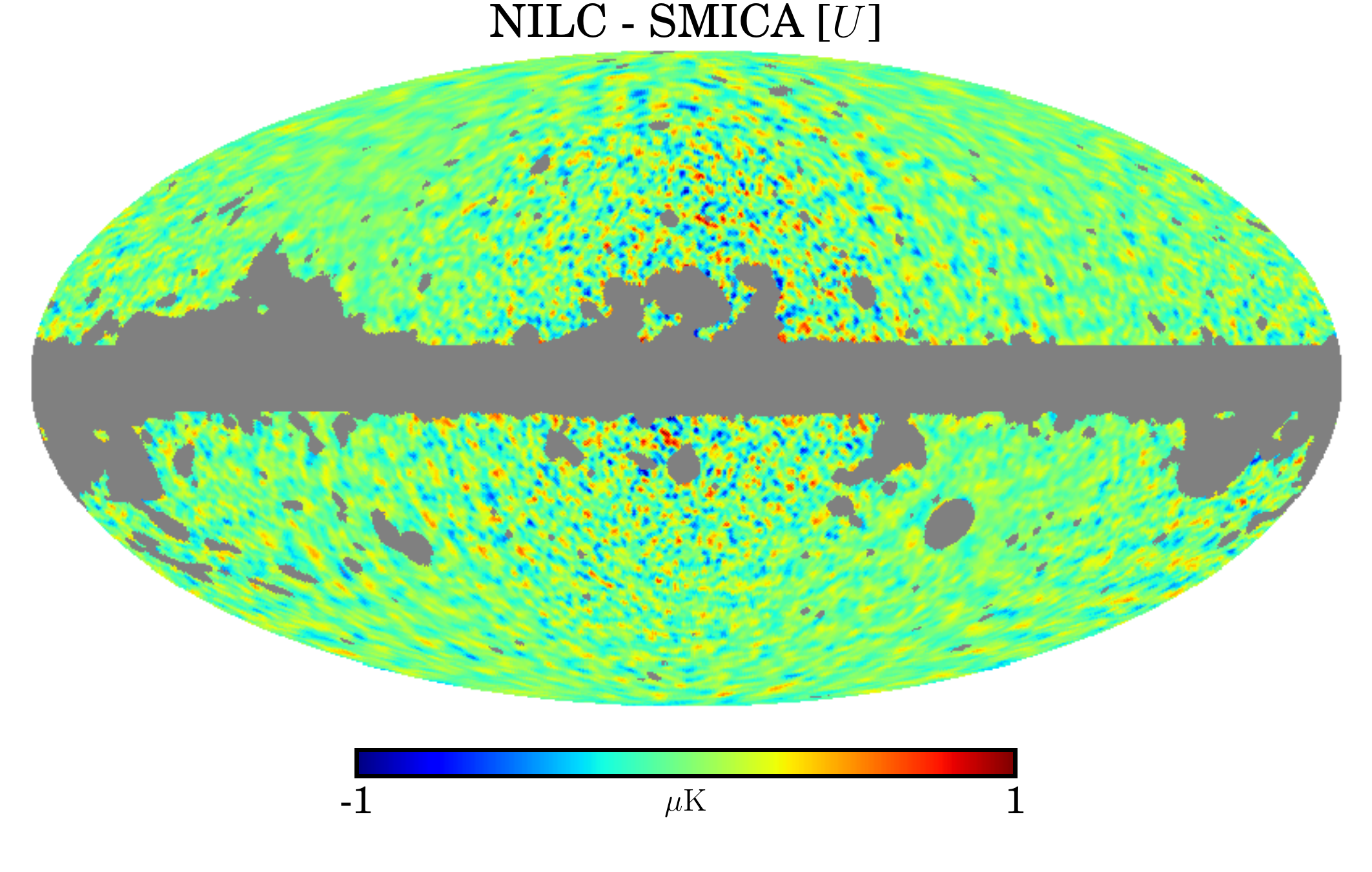}
\end{tabular}
\caption{Planck \textit{data.} Differences between the axisymmetric limit \((N = 1)\) of Spin-SILC, NILC and SMICA. The maps have been smoothed to \(\mathrm{FWHM} = 80\arcmin\) and downgraded to \(N_\mathrm{side} = 128\). The grey pixels are the UPB77 confidence mask from \citet{2015arXiv150205956P}, which masks the regions of the NILC and SMICA maps not recommended for cosmological analysis. The differences are (\textit{from top to bottom}) (a) SILC \((N = 1)\) - NILC, (b) SILC \((N = 1)\) - SMICA and (c) NILC - SMICA; and in (\textit{from left to right}) (i) Stokes \(Q\) and (ii) Stokes \(U\) maps.}
\label{fig:nilcdiffMap}
\end{figure*}

\begin{figure}
\includegraphics[width=\columnwidth]{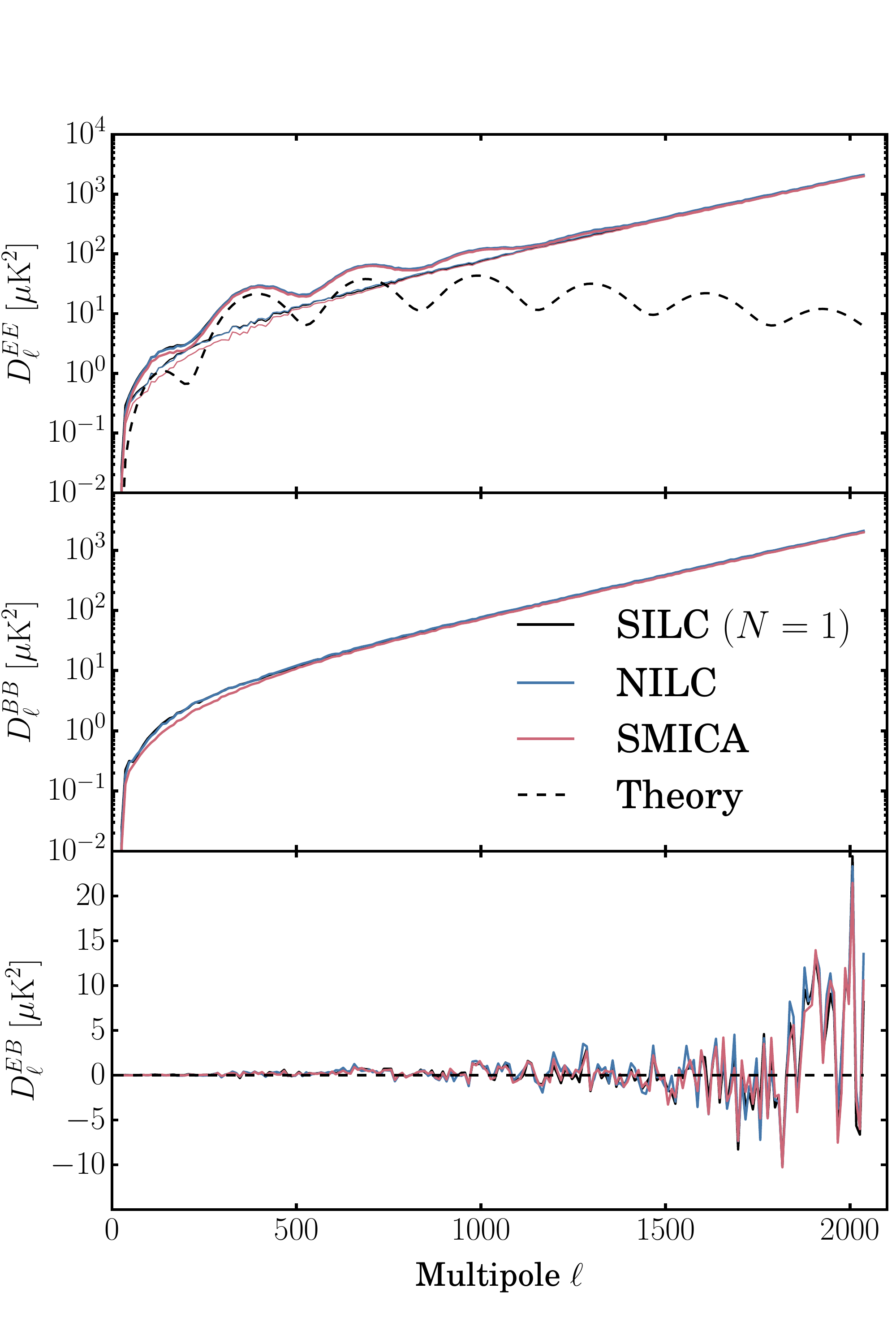}
\caption{Planck \textit{data.} \textit{From top to bottom,} (a) \(EE\), (b) \(BB\) and (c) \(EB\) angular power spectra comparing the axisymmetric limit \((N = 1)\) of Spin-SILC to NILC and SMICA. \textit{In the top panel} (a), the thin lines show residuals after subtracting the best-fit \(\Lambda\)CDM model from the {\Planck} 2015 likelihood.}
\label{fig:nilcspec}
\end{figure}

Having presented the main results of applying Spin-SILC to {\Planck} data, we can perform a validation check by comparing to other component separation reconstructions of the CMB from the same dataset. For this purpose, we concentrate on the methods NILC \citep{2009A&A...493..835D} and SMICA \citep{2008arXiv0803.1814C}, which are two of the four methods used internally by the Planck Collaboration \citep{2015arXiv150205956P}: the former because it is the most similar method to Spin-SILC and the latter because it is the baseline method adopted by the Planck Collaboration for high-resolution analyses. Like Spin-SILC, NILC is an internal linear combination (ILC) method performed in wavelet space. Unlike Spin-SILC, NILC uses scalar axisymmetric wavelets, specifically scalar needlets \citep{2006SIAM...38...574,marinucci:2008,baldi:2009}, rather than the spin directional wavelets we use \citep{mcewen:s2let_spin, 2015arXiv150203120L, 2014IAUS..306...64M} (although spin needlets \citep{geller:2008} and mixed needlets \citep{geller:2010} have also been developed). This means that in its extension to polarisation \citep{2013MNRAS.435...18B,2015arXiv150205956P}, NILC acts independently on input \(E\) and \(B\) maps, having been previously decomposed from the original Stokes \(Q\) and \(U\) data. Similarly to ILC methods, SMICA forms a linear combination of multifrequency data, but in harmonic space. Unlike blind ILC methods which require no physical modelling of the sky components, SMICA is only semi-blind in that on large scales, rather than empirically estimating covariances on the data (as on small scales), a fit is performed to a model of the component covariances, with the option to constrain these covariances. This is extended to polarisation by performing a joint processing of the \(E\) and \(B\) modes in harmonic space. A further difference between ILC methods and SMICA is that SMICA has no spatial localisation in its component separation, although a wavelet implementation of SMICA does exist \citep{2005EJASP2005..100M}. Indeed, Spin-SILC localises with regard to the greatest number of domains of information, allowing spatial, harmonic and morphological localisation through the use of directional wavelets (see \citealt{2016arXiv160101322R} for a discussion of the morphological localisation properties of directional wavelets in the SILC method). A significant advantage of Spin-SILC over existing component separation methods (including NILC and SMICA) is the use of spin wavelets, which allows simultaneous component separation and \(E\)-\(B\) decomposition.

We can empirically compare the three methods with an analysis of the CMB maps reconstructed from the (full-mission 2015 release) {\Planck} data and full-sky power spectra measured from those maps. Figure~\ref{fig:nilcdiffMap} shows the differences between the CMB reconstructed by Spin-SILC (in the axisymmetric limit \(N = 1\)), NILC and SMICA. We show differences in \(Q\) and \(U\) maps as this most directly compares to the map products provided by the Planck Collaboration. The differences between the three methods are small in magnitude in both \(Q\) and \(U\) and mostly concentrated at the edges of the Galactic mask towards the Galactic centre, where foreground emission is most intense and complex. Quantitatively, we can compare the mean values and standard deviations of the full-sky difference maps. The mean values of the \(Q\) difference maps in Figs.~\ref{fig:nilcdiffMap} (i) (a), (b) and (c) (\textit{from top to bottom on the left-hand side}) are respectively \(5.2 \times 10^{-5}\), \(4.3 \times 10^{-5}\) and \(-9.3 \times 10^{-6}\) \(\mu \mathrm{K}\), while the standard deviations are 0.34, 0.37 and 0.34 \({\mu \mathrm{K}}^2\). The mean values of the \(U\) difference maps in Figs.~\ref{fig:nilcdiffMap} (ii) (a), (b) and (c) (\textit{from top to bottom on the right-hand side}) are respectively \(6.4 \times 10^{-5}\), \(-6.9 \times 10^{-5}\) and \(-1.3 \times 10^{-4}\) \(\mu \mathrm{K}\), while the standard deviations are 0.33, 0.37 and 0.31 \({\mu \mathrm{K}}^2\). These values are small and similar, suggesting a strong consistency between the three methods. As discussed in \S~\ref{sec:input}, some of the input {\Planck} data are high-pass filtered and so are also the output results of all three methods (for \(\ell < 40\)). This means that any comparison can only be carried out for \(\ell \geq 40\).

Figure~\ref{fig:nilcspec} compares full-sky power spectra measured from component separation maps with CMB spectra derived from the {\Planck} 2015 \(TT\) and low \(TEB\) likelihood\footnote{The parameters come from the \texttt{base\_plikHM\_TT\_lowTEB} likelihood. The values are available in the {\Planck} 2015 Release Explanatory Supplement: 2015 Cosmological parameters and MC chains (\url{http://wiki.cosmos.esa.int/planckpla2015/images/f/f7/Baseline_params_table_2015_limit68.pdf}).}. The three panels respectively compare \(EE\), \(BB\) and \(EB\) spectra. As with the simulated results in \S~\ref{sec:simulations}, the \(EE\) and \(BB\) spectra from all three methods are consistent with significant residual noise power due to the noisiness of the input maps. The only discernible difference is marginally less power in the SMICA maps at multipoles around \(\ell = 250\). This could be attributed to the semi-blindness of SMICA better characterising the noise properties of the {\Planck} data. The two blind methods, Spin-SILC and NILC have near-identical spectra at all multipoles. The \(EB\) spectra of all three methods are consistent with zero.

The comparison of Spin-SILC to existing methods NILC and SMICA has strongly validated the results we showed in \S~\ref{sec:data}. An analysis of maps and power spectra shows an internal consistency between the three algorithms. It also shows that the power of Spin-SILC and other component separation methods is limited by the low S/N of the {\Planck} polarisation data, with large amounts of residual noise in the reconstructed CMB. The full potential of the Spin-SILC method thus awaits the input of higher-S/N polarisation data available from upcoming CMB observations.

\section{Discussion}
\label{sec:discussion}

The testing of Spin-SILC on {\Planck} simulations in \S~\ref{sec:simulations} and data in \S~\ref{sec:data} shows that the use of spin wavelets in CMB polarisation component separation can successfully reconstruct the cosmological background. This is particularly true of the \(EE\) power spectrum with the clear detection of the first four acoustic peaks in both simulations and real data. The residual maps to the input simulated CMB (Fig.~\ref{fig:ffp8N1diffMap}) and power spectra estimated from the SILC maps (Fig.~\ref{fig:ffp8N1spec}) show high levels of residual noise, reflecting the relatively low S/N of the {\Planck} data we used. In \S~\ref{sec:comp}, we carried out a comparison of the Spin-SILC method with two of the most accurate existing component separation algorithms, NILC and SMICA. We validated our main results by showing a strong internal consistency in reconstructed CMB maps (Fig.~\ref{fig:nilcdiffMap}) and power spectra measured from those maps (Fig.~\ref{fig:nilcspec}). However, this comparison also revealed high levels of residual noise in the CMB estimated by all three methods, due to the S/N limitations of the {\Planck} polarisation data.

We also tested the use of directional spin wavelets in the Spin-SILC method on the simulations in \S~\ref{sec:simulations}. We found very modest reductions in reconstructed residual power (Fig.~\ref{fig:ffp8N10spec}) as the amount of directionality was increased. However, given the low S/N input data, the magnitude of these reductions is much smaller than the overall amount of residual power, though at the accuracy required to reconstruct the cosmological \(BB\) signal, this level of power reduction may become relevant in the high S/N regime. As discussed in \citet{2016arXiv160101322R} in the scalar SILC method, instrumental noise has no particular directional structure and thus in the low S/N regime the use of directionality is expected to only have a small effect on the estimate of the reconstructed CMB. The community is only just beginning to accurately characterise polarised foregrounds at high resolution at a range of frequencies. If the foregrounds are complex in high S/N observations, the ability to use directional wavelets may prove useful in localising component separation according to the morphology of the CMB and foregrounds.

Spin-SILC introduces a number of novelties into CMB polarisation component separation. Most notably, the use of spin wavelets allows simultaneous \(E\)-\(B\) decomposition and the joint minimisation of \(E\) and \(B\) auto-correlations in residual contamination. Moreover, the use of directionality allows the fine-tuning of the cleaning algorithm according to the morphology of the local signal. As discussed in \citet{2016arXiv160101322R}, there are various sources of error in the ILC method, which will affect the spin-ILC in an equivalent fashion. Of particular note is the ILC bias, corresponding to the empirical cancellation of CMB modes due to chance correlations with foregrounds and noise (see \citealt{2009A&A...493..835D} for a fuller discussion of this effect), which will also affect the reconstruction of the CMB polarisation. The amount of cancellation may increase with the amount of directionality used within the method. In \citet{2016arXiv160101322R}, we showed that this can be mitigated either directly from the data or through suites of simulations.

\section{Conclusions}
\label{sec:concs}

We have presented Spin-SILC, a foreground component separation method specifically developed for the analysis of CMB polarisation data. The use of spin wavelets allows the full analysis of the spin-2 polarisation signal \(P = Q + iU\), formed by the Stokes \(Q\) and \(U\) parameters. By the particular construction of the spin wavelets we use, Spin-SILC carries out the decomposition of the polarisation signal into \(E\) and \(B\) modes by separating the real and imaginary parts of the complex spin-2 wavelet coefficients. This occurs simultaneously to the component separation, where the auto-correlations of \(E\) and \(B\) modes are jointly minimised in residual contamination to the reconstructed CMB. Moreover, the wavelets we use are directional. This allows different directional morphologies of CMB and polarised foreground to be separated. This extra information can then be used to better localise the Spin-SILC cleaning algorithm.

We have tested Spin-SILC on full-mission {\Planck} simulations and data. We showed that the method can accurately extract cosmological information from input \(Q\) and \(U\) maps. We also validated our main results with a comparison to the internal {\Planck} methods, NILC and SMICA, showing a strong consistency in both CMB maps and power spectra, with small residuals compared to the two. However, we note that the analysis in this paper is limited by the low S/N of the {\Planck} polarisation data. Our final \(E\) and \(B\) maps (as well as those of NILC and SMICA) are dominated by residual instrumental noise. Moreover, the full power of the use of directionality in Spin-SILC cannot be fully explored due to the high level of noise in the {\Planck} input data. If polarised foregrounds have complex morphology in the high S/N regime, then the use of directionality may prove a useful extra tool in extracting the CMB. In general, it will be interesting to test Spin-SILC further with the high S/N data of upcoming CMB polarisation observations. We make our \(Q\), \(U\), \(E\) and \(B\) maps available at \url{http://www.silc-cmb.org}\footnote{The DOI for our data release is 10.5281/zenodo.50579.}.

Furthermore, Spin-SILC can be combined with the estimators of \citet{ebsep} to perform component separation on the cut-sky and give accurate estimates of pure \(E\) and \(B\) modes (pure \(E\) (\(B\)) modes are orthogonal to all \(B\) (\(E\)) modes on the cut-sky, respectively). It achieves this in a straightforward fashion (with only two additional wavelet transforms of the input data) due to the construction of the spin wavelets (see \citealt{ebsep} for more details about pure mode estimation on the cut-sky using spin wavelets). This is of particular importance for the upcoming high resolution, high S/N CMB polarisation experiments, which will typically make partial-sky observations. Spin-SILC will provide a computationally-efficient algorithm to perform simultaneous \(E\)-\(B\) decomposition and accurate foreground component separation for these next-generation experiments.

\section*{Acknowledgements}
\label{sec:ack}

KKR thanks Franz Elsner and Stephen Feeney for valuable discussions. KKR was supported by the Science and Technology Facilities Council. HVP and BL were partially supported by the European Research Council under the European Community's Seventh Framework Programme (FP7/2007-2013) / ERC grant agreement number 306478-CosmicDawn. JDM was partially supported by the Engineering and Physical Sciences Research Council (grant number EP/M011852/1). AP was supported by the Royal Society. Based on observations obtained with {\Planck} (\url{http://www.esa.int/planck}), an ESA science mission with instruments and contributions directly funded by ESA Member States, NASA and Canada.

\bibliographystyle{mymnras_eprint}
\bibliography{bib_journal_names_short,spin_silc_mnras}

\end{document}